\documentclass[aps,pre,onecolumn,floatfix]{revtex4}

\usepackage{bm}
\usepackage{graphicx}
\usepackage{amssymb,amsfonts,amsmath}
\usepackage{epsf}
\usepackage{color}
\usepackage{hyperref}
\usepackage{bbold}
\usepackage{subfigure}
\usepackage{epstopdf}
\usepackage{soul}
\usepackage{mathrsfs}

\usepackage{upgreek}

\newcommand{\be}{\begin{equation}}
\newcommand{\ee}{\end{equation}}
\newcommand{\bea}{\begin{eqnarray}}
\newcommand{\eea}{\end{eqnarray}}

\begin{document}

\title{\bf Hydrodynamic properties of the perfect hard-sphere crystal: \\ Microscopic computations with Helfand moments}

\author{Jo\"el Mabillard}
\email{Joel.Mabillard@ulb.be; \vfill\break ORCID: 0000-0001-6810-3709.}
\author{Pierre Gaspard}
\email{Gaspard.Pierre@ulb.be; \vfill\break ORCID: 0000-0003-3804-2110.}
\affiliation{Center for Nonlinear Phenomena and Complex Systems, Universit{\'e} Libre de Bruxelles (U.L.B.), Code Postal 231, Campus Plaine, B-1050 Brussels, Belgium}

\begin{abstract}
\noindent{\bf Abstract}: Within the framework of the local-equilibrium approach, the equilibrium and nonequilibrium properties relevant to the hydrodynamics of the perfect hard-sphere crystal are obtained with molecular dynamics simulations using the Helfand moments associated with momentum and energy transports.  Since this crystal is face-centered cubic, the hydrodynamic properties we consider are the hydrostatic pressure, the isothermal bulk modulus, the specific heat capacities and their ratio, the three isothermal elastic constants $(C_{11}^T,C_{12}^T,C_{44}^T)$, the heat conductivity, and the three viscosities $(\eta_{11},\eta_{12},\eta_{44})$ (in Voigt's notations).  These properties are computed as a function of the particle density.  The pressure and the transport coefficients diverge near the close-packing density, as the collision frequency per particle does.
\end{abstract}

\maketitle

\section{Introduction}
\label{sec:Introduction}

Understanding the macroscopic hydrodynamics of crystals in terms of the microscopic motion of their atoms is one of the major challenges in contemporary statistical mechanics~\cite{W65,W67,W98,GAW70,BA71,MPP72,F75,FC76,L88,PBHB20,KDEP90,SE93,S97,FL87,PF03,KK04,L20,MG20,MG21,DvBK21,H22,MGHF22,H23}.  In contrast to fluids that have isotropic and uniform properties, crystals are characterized by long-range periodic order in the three-dimensional space.  As a consequence, the continuous symmetries under the group of spatial translations and rotations prevailing in fluids are broken into the discrete symmetries of one among the 230 space groups, which are known for the classification of crystal structures~\cite{AM76}.  By the Nambu-Goldstone mechanism~\cite{N60,G61}, the breaking of the three continuous symmetries under spatial translations generates three extra slow modes in addition to the five slow modes expected from the five fundamental conservation laws of energy, momentum, and mass in one-component crystals. Therefore, their macroscopic hydrodynamics is ruled by eight slow modes, including three pairs of propagating sound modes in the longitudinal and the two transverse directions, the heat mode, and the vacancy diffusion mode. 

Often, the vacancy diffusion mode is much slower than the seven other slow modes and can thus be neglected in a first approximation to the macroscopic description of the crystal.  In this case, the crystal is said to be {\it perfect}.  In perfect crystals, the hydrodynamic properties include the equilibrium equations of state for energy and pressure, the equilibrium properties of elasticity, and the nonequilibrium transport properties, which are the viscosities and the heat conductivities.  The elastic constants give the propagation velocities of the six sound modes, and the transport coefficients determine the damping rates of all the modes.

All these properties should be quantitatively known in order to establish the macroscopic hydrodynamic equations of the crystal and to predict the analytic form of the correlation and spectral functions characterizing the fluctuations of the perfect crystal. Although there exist many results reported on the equilibrium properties of crystals, their nonequilibrium properties are only partially known.  In particular, several papers have been devoted to heat conduction in the hard-sphere crystal \cite{GAW70,PBHB20}, but their viscosities do not seem to have been systematically investigated to our knowledge. In this context, a challenging issue of nonequilibrium statistical mechanics is to obtain all the transport properties of crystals, including their viscosities, in terms of the microscopic dynamics with the same level of reliance as in fluids.

\begin{figure}[h!] \centering
    \begin{minipage}{0.5\textwidth}
        \centering
        \includegraphics[width=0.6\textwidth]{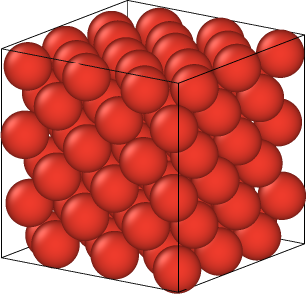} 
    \end{minipage}\hfill
    \begin{minipage}{0.5\textwidth}
        \centering
        \includegraphics[width=0.6\textwidth]{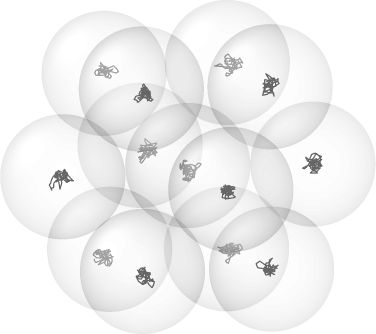} 
    \end{minipage}
\caption[] {Left: Crystal of $N=108$ hard spheres at  the density $n_*=1.2$.  The solid black lines show the simulation box. Right: Trajectories of the particles during the time interval $\Delta t_*=1$, showing one particle in the crystal and its twelve nearest neighbors. The figures have been generated with OVITO~\cite{S10}.}\label{Fig:Crystal}
\end{figure}

The purpose of this paper and the following two~\cite{MG23c,MG23d} is to perform for perfect crystals the same study as we have previously carried out for fluids~\cite{MG23b} and to test the predictions of the local-equilibrium approach to the hydrodynamics of crystals~\cite{MG20,MG21}.  As shown within the framework of the local-equilibrium approach~\cite{MG23a}, the equilibrium and nonequilibrium properties of crystals can be computed by using respectively the mean values and the covariances of the so-called Helfand moments \cite{H60}.  As in reference~\cite{MG23b}, the vehicle of our study is the hard-sphere system, now at high densities, where the system becomes solid, forming a face-centered cubic (fcc) crystal.  Figure~\ref{Fig:Crystal} shows the crystal and trajectories of some of its hard spheres in numerical simulations.  The crystal is perfect since there is a hard sphere in the vicinity of all the sites of the fcc lattice.  The motion of the hard spheres has small amplitudes around the lattice sites, as observed in the figure.

Since the interaction between the hard spheres is only repulsive, the crystal exists because of the external pressure due to the periodic boundary conditions implemented for simulating the microscopic dynamics.  Accordingly, the elasticity properties that are relevant for such crystals are the strain-stress coefficients \cite{W98}, rather than the elastic constants used for crystals holding together by both attractive and repulsive interatomic forces without necessitating an additional external pressure.  The strain-stress coefficients are related to the elastic constants and the hydrostatic pressure \cite{W98}.

The hard-sphere system is also interesting because its dynamics is collisional and, thus, significantly different from the dynamics of normal modes usually considered in solid-state theory \cite{AM76}.  Therefore, the macroscopic hydrodynamics of the hard-sphere system should be directly formulated using the fundamental conservation laws combined with the Nambu-Goldstone mechanism without the help of the standard harmonic model for crystals.

In this paper, our aim is thus to compute all the coefficients needed for the hydrodynamics of the perfect hard-sphere crystal, i.e., the hard-sphere crystal without vacancy.  For this goal, the numerical simulations are carried out for a completely filled fcc crystal of hard spheres.  For this perfect crystal, the Helfand moments to consider are  those associated with the transports of momentum and energy.  The hydrostatic pressure and the elastic constants are obtained with the mean values of the Helfand moments, and the viscosities and heat conductivities with their covariances.  Since the hard-sphere crystal is cubic, there are three elastic constants, three viscosity coefficients, and one heat conductivity to compute.

The plan of this paper is the following.  In section~\ref{sec:statmech}, the statistical mechanics of   one-component crystals is presented with the focus on the case of perfect crystals.  Their Helfand moments are defined and used to obtain the equilibrium and nonequilibrium properties relevant to hydrodynamics.  In section~\ref{sec:Results_HS}, the general methods are applied to obtain the hydrostatic pressure, the isothermal bulk modulus, the specific heat capacity at constant pressure, the specific heat ratio, the three elastic constants, the three viscosities, and the heat conductivity as a function of the particle density for the perfect hard-sphere crystal.

{\it Notations.} The Latin indices $a, b, c, \ldots = x, y, z$ correspond to spatial coordinates and the Greek indices $\alpha, \beta, \ldots$ label the hydrodynamic variables. The indices $i,j,\ldots = 1,2,\ldots$ label the particles. Unless explicitly stated, Einstein's convention of summation over repeated indices is adopted. $h$ denotes Planck's constant, $k_{\rm B}$ Boltzmann's constant, and ${{\rm i}}=\sqrt{-1}$.

\section{Statistical mechanics of one-component crystals}
\label{sec:statmech}

We consider a crystal composed of $N$ atoms of mass $m$. The positions ${\bf r}_i=(r_i^a)$ and the momenta ${\bf p}_i=(p_i^a)$ of the atoms form the phase-space variables $\Gamma=({\bf r}_i,{\bf p}_i)_{i=1}^N\in{\mathbb R}^{6N}$ of the system. The energy potential of the mutual interaction between the atoms is denoted ${\cal V}(r_{ij})$, where $r_{ij}=\Vert{\bf r}_i-{\bf r}_j\Vert$ is the distance separating the $i^{\rm th}$ and $j^{\rm th}$ atoms.  The total energy of the crystal is thus given by the Hamiltonian function
\begin{align}\label{Hamilt_fn}
H = \sum_i \frac{{\bf p}_i^2}{2m} + \frac{1}{2}\sum_{i\neq j} {\cal V}(r_{ij}) \, .
\end{align}
Consequently, the dynamics of the atoms is ruled by Hamilton's equations
\begin{align}\label{Hamilt_eqs}
\frac{{\rm d}r_i^a}{{\rm d}t} = \frac{\partial H}{\partial p_i^a} = \frac{p_i^a}{m} \, , \qquad\qquad
\frac{{\rm d}p_i^a}{{\rm d}t} = -\frac{\partial H}{\partial r_i^a} = \sum_{j(\ne i)} F_{ij}^a \, ,
\end{align}
where $F_{ij}^a \equiv - {\partial {\cal V}(r_{ij})}/{\partial r_i^a}$ is the interaction force exerted on the $i^{\rm th}$ atom by the $j^{\rm th}$ atom.  

To simulate the properties of bulk matter, the $N$ atoms should evolve in a spatial domain of finite volume.  To compute the hydrostatic pressure and the transport properties, we take a cubic domain with periodic boundary conditions, thus forming a torus.  Accordingly, the Cartesian coordinates of the positions should satisfy
\be
0 \leq r_i^a < L \qquad\mbox{for all}\qquad a=x,y,z \quad\mbox{and}\quad i=1,2,\dots, N \, ,
\label{eq:std_pbc}
\ee
where $L$ is the length of the sides of the cube, so that the volume of this torus is equal to $V=L^3$.

For the purpose of computing the elastic properties, this cubic domain is deformed by the transformation mapping each point of the cube, ${\bf R}=(R^a)\in [0,L[^3$, onto the new point ${\bf r}=(r^a)$ given by $r^a=D^{ab} R^b$ in terms of the deformation matrix $\boldsymbol{\mathsf D}=(D^{ab})$.  By this deformation, the cubic domain may be contracted, stretched, or sheared in different directions.  After deformation, the periodic boundary conditions are expressed as
\be
\sum_{b(\ne a)} D^{ab} r_i^b \leq r_i^a < D^{aa} L + \sum_{b(\ne a)} D^{ab} r_i^b \qquad\mbox{for all}\qquad a=x,y,z \quad\mbox{and}\quad i=1,2,\dots, N \, .
\label{eq:deformed_pbc}
\ee
The volume of the deformed cube has the value $\tilde V = L^3 \det\boldsymbol{\mathsf D}$.  For the properties of linear elasticity, the deformations can be taken to be arbitrarily small such that $\Vert \boldsymbol{\mathsf D}^{\rm T}\cdot \boldsymbol{\mathsf D} - \boldsymbol{\mathsf 1}\Vert \ll 1$, where $\boldsymbol{\mathsf 1}$ is the identity matrix.

We also assume that the interaction force has a finite range smaller than half the size of the spatial domain $L/2$.

Since the Hamiltonian function does not depend on time, $\partial_tH=0$, and only depends on the distances between the atoms, this Hamiltonian dynamics is invariant under continuous spatiotemporal translations, which leads to the conservation of the total energy $E=H$ and the total momentum ${\bf P}=\sum_i{\bf p}_i$.  Moreover, the total mass $M=mN$ is also conserved.  Therefore, molecular dynamics are carried out in the $(N,V,E)$-ensemble, where $N$, $V$, and $E$ take fixed values, and the total momentum is taken equal to zero ${\bf P}=0$.

\subsection{Local conservation laws and symmetry breaking}
\label{sec:conserve}

In  one-component crystalline solids, the relaxation towards equilibrium is controlled by the eight hydrodynamic modes~\cite{MPP72,F75,FC76}.  They are the slowest modes of the dynamics because they are characterized by dispersion relations $\omega({\bf q})$ vanishing with  their wave number $q=\Vert{\bf q}\Vert$. These eight modes originate from the spontaneous breaking of the translational symmetry in the three spatial directions and from the five fundamental conservation laws of energy, momentum, and mass.  

Indeed, the microscopic densities $\hat{c}^{\alpha}({\bf r};\Gamma)$ of these five quantities obey local conservation laws of the following form,
\be
\partial_t \, \hat c^{\alpha} + \nabla^a \hat J_{c^\alpha}^a = 0 \, ,
\label{eq:local-eqs}
\ee
where $\hat J_{c^\alpha}^a({\bf r};\Gamma)$ denotes the corresponding current densities, as can be derived from the microscopic  Hamiltonian dynamics.  The microscopic densities of energy, momentum, and mass are respectively defined as $\hat\epsilon({\bf r};\Gamma) \equiv \sum_i \varepsilon_i \delta({\bf r}-{\bf r}_i)$ with $\varepsilon_i \equiv {{\bf p}_i^2}/{2m} + \frac{1}{2} \sum_{j(\ne i)} {\cal V}(r_{ij})$, $\hat{g}^a({\bf r};\Gamma) \equiv \sum_i p_i^a \delta({\bf r}-{\bf r}_i)$, and $\hat\rho =m \hat n$ with the particle density $\hat n({\bf r};\Gamma) \equiv \sum_i \delta({\bf r}-{\bf r}_i)$.

Furthermore, the crystalline solid has long-range order in the form of a spatially periodic equilibrium mean particle density, $n_{\rm eq}({\bf r})=\langle \hat n({\bf r};\Gamma)\rangle_{\rm eq}$ \cite{MG21}.  Therefore, the equilibrium properties of the crystal are no longer invariant under the continuous group of three-dimensional spatial translations, but under one of the 230 discrete  space groups.  According to the Nambu-Goldstone mechanism, the phenomenon of continuous symmetry breaking implies that the hydrodynamics of the crystal should involve not only the densities of the five fundamentally conserved quantities, but also the three local order parameters associated with the broken continuous symmetries, namely, the three components of the displacement field.  Remarkably, a microscopic expression for the displacement field has been identified in references~\cite{SE93,S97} as
\begin{align}
\label{eq:displu}
\hat{u}^a({\bf r};\Gamma) = -\, (\pmb{\cal N}^{-1})^{ab} \int {\rm d}{\bf r'}  \,\int_{\cal B} \frac{{\rm d}{\bf k}}{(2\pi)^3} \, {\rm e}^{{\rm i} {\bf k}\cdot({\bf r}-{\bf r'})} \, {\nabla'}^b n_{\rm eq}({\bf r'}) \, \left[\hat n({\bf r'};\Gamma)-n_{\rm eq}({\bf r'})\right] ,
\end{align}
${\cal B}$ denoting the first Brillouin zone of the crystalline lattice and $\pmb{\cal N}=({\cal N}^{ab})$  given by
\begin{align}\label{N-dfn}
{\cal N}^{ab} \equiv \frac{1}{v} \int_{v} {\rm d}{\bf r}\ \nabla^a n_{\rm eq}({\bf r}) \, \nabla^b n_{\rm eq}({\bf r}) \, ,
\end{align}
where $v$ is the volume of a primitive unit cell of the lattice. 
Its time derivative $\partial_t\hat u^{a}=-\hat{J}_{u^a}$ can be deduced from the microscopic Hamiltonian dynamics.  Under conditions close to equilibrium, where the dynamics of the crystal is ruled by the properties of linear elasticity, we may consider the linear strain tensor, $\hat{u}^{ab} \equiv \frac{1}{2} \big(\nabla^a\hat{u}^b +\nabla^b\hat{u}^a\big)$, which also obeys a local conservation equation of the form~\eqref{eq:local-eqs} with the current densities $\hat{J}^a_{u^{bc}}=\frac{1}{2} \big( \delta^{ab} \hat{J}_{u^c}+\delta^{ac}\hat{J}_{u^b}\big)$.

Therefore, the microscopic hydrodynamics of the crystal can be expressed in terms of the local conservation equations~\eqref{eq:local-eqs} with the densities $\hat c^{\alpha}=(\hat\rho,\hat\epsilon,\hat g^b, \hat{u}^{bc})$ and the corresponding current densities $\hat J_{c^\alpha}^a=\big(\hat J_{\rho}^a,\hat J_{\epsilon}^a,\hat J_{g^b}^a,\hat J_{u^{bc}}^a\big)$. Now, the basic issue is to deduce the equations of macroscopic hydrodynamics from the microscopic Hamiltonian description, including the dissipative effects and microscopic formulas for the transport coefficients. Several approaches for the crystal have been introduced~\cite{S97,H22,MGHF22,H23}. In the following, we consider the local-equilibrium approach~\cite{MG20,MG21,MG23a}.

\subsection{The local-equilibrium approach}
\label{sec:localeq}

In principle, the macroscopic equations can be obtained  by taking the statistical averages $\langle\cdot\rangle_t\equiv\int {\rm d}\Gamma\, {\cal P}_t(\Gamma)\, (\cdot)$ of the local conservation equations~\eqref{eq:local-eqs} with respect to the probability distribution ${\cal P}_t$ ruled by the Liouville equation, giving
\begin{align}
\partial_t\,  \langle\hat{c}^\alpha\rangle_t  + {\nabla}^a \langle\hat{J}^a_{c^\alpha}\rangle_t   = 0 \, .
\end{align}
The mean values of the microscopic densities are thus expected to define the macroscopic densities, $c^\alpha({\bf r},t)\equiv\langle\hat{c}^\alpha\rangle_t$.  

In parallel with this exact time evolution, the local-equilibrium approach considers the coarse-grained local Gibbsian probability distribution
\begin{align}
{\cal P}_{{\rm leq},{\boldsymbol{\lambda}_t}}(\Gamma) = \frac{1}{\Delta\Gamma}\, {\rm e}^{- \int {\rm d}{\bf r} \, {\lambda}^\alpha({\bf r},t)\,\hat{c}^\alpha({\bf r};\Gamma)- \Omega(\boldsymbol{\lambda}_t) } \, ,
\label{eq:leq_distrib}
\end{align}
where $ \Delta\Gamma\equiv h^{3N}$, the so-called Massieu functional $\Omega$ guarantees the normalization of the probability distribution, and $\boldsymbol{\lambda}_t=[\lambda^{\alpha}({\bf r},t)]$ are fields conjugated to the densities.  These conjugate fields are such that the mean values of the densities with respect to the exact distribution ${\cal P}_t$ are equal to their mean values with respect to the local-equilibrium distribution~\eqref{eq:leq_distrib}: $\langle \hat{c}^\alpha\rangle_t=\langle\hat{c}^\alpha\rangle_{{\rm leq},\boldsymbol{\lambda}_t}$.  These conditions should hold at every position ${\bf r}$ in the system and every time~$t$ of the time evolution and they play a central role in the local-equilibrium approach, in particular, because they determine the values of the conjugate fields $\boldsymbol{\lambda}_t$ in terms of the mean values of the densities $\langle\hat{c}^\alpha\rangle_t$.  

The link to the thermodynamics of crystals can be established within this approach \cite{MG20,MG21}.
The conjugate fields corresponding to the densities $\hat c^{\alpha}=(\hat\rho,\hat\epsilon,\hat g^b, \hat{u}^{bc})$ can be expressed up to terms of second order in the gradients as $\lambda^\alpha=(\lambda_\rho=-\beta\mu,\lambda_\epsilon=\beta,\lambda_{g^b}=-\beta v^b,\lambda_{u^{bc}}=-\beta\phi^{bc})$, where $\beta=(k_{\rm B}T)^{-1}$ denotes the inverse temperature, $\mu$ the chemical potential, $v^b$ the velocity field, and $\phi^{bc}$ a tensor describing the stress applied to the crystal by perturbations.

Without much restriction, the initial probability distribution  may be taken as the local-equilibrium distribution corresponding to the initial values of the macrofields $\boldsymbol{\lambda}_0$ in the system, ${\cal P}_0={\cal P}_{{\rm leq},\boldsymbol{\lambda}_0}$.  Hence, the exact probability distribution at time $t$ can be expressed as ${\cal P}_t(\Gamma) = {\rm e}^{-\widehat{\mathcal L} t} \, {\cal P}_{{\rm leq},{\boldsymbol{\lambda}_0}}(\Gamma)$, where $\widehat{\mathcal L}$ is the Liouvillian operator defined as $\widehat{\mathcal L}A \equiv \{A,H\}$ in terms of the Poisson bracket with the Hamiltonian.  However, time evolution generates more and more pronounced differences between the exact probability distribution ${\cal P}_t$ and the local-equilibrium distribution~\eqref{eq:leq_distrib} as time increases.  These differences can be captured with the quantity \cite{McL63,S14}
\begin{align}
\Sigma_t(\Gamma)  \equiv   \int_0^t {\rm d}\tau\ \partial_\tau \Big[ \int {\rm d}{\bf r} \, {\lambda}^{\alpha}({\bf r},\tau)\, \hat{c}^\alpha({\bf r};\Gamma_{\tau - t})+ \Omega(\boldsymbol{\lambda}_\tau)\Big] ,
 \end{align}
which is introduced to satisfy the following equality at every time $t$,
\begin{align}
{\cal P}_t(\Gamma)  =   {\rm e}^{\Sigma_t(\Gamma)}\, {\cal P}_{{\rm leq},\boldsymbol{\lambda}_t}(\Gamma)\, .
\end{align}
As a consequence, the mean value of any observable quantity $A(\Gamma)$ with respect to the exact probability distribution can be expressed as $\langle {A}\rangle_t  = \langle {A}\, {\rm e}^{\Sigma_t}\rangle_{{\rm leq},{\boldsymbol{\lambda}}_t}$ in terms of the local-equilibrium distribution.  This approach allows us to decompose the mean values of the current densities into their reversible and dissipative parts, because the former conserves entropy although the latter does not \cite{McL63,S14,MG23a,MG20,MG21}.  In this way, the macroscopic equations of hydrodynamics can be written as
\begin{align} 
\partial_t\, {c}^\alpha(\mathbf{r},t)  +  {\nabla}^a \big[{J}^{{\rm R},a}_{c^\alpha}(\mathbf{r},t)  +{J}^{{\rm D},a}_{c^\alpha}(\mathbf{r},t)  \big] & = 0\;,
\end{align}
where the reversible part of the current densities is defined as ${J}^{{\rm R},a}_{c^\alpha}(\mathbf{r},t) \equiv \langle\hat{J}^a_{c^\alpha}(\mathbf{r};\Gamma)\rangle_{{\rm leq},\boldsymbol{\lambda}_t}$ and its dissipative part as ${J}^{{\rm D},a}_{c^\alpha}(\mathbf{r},t) \equiv \langle\hat{J}^a_{c^\alpha}(\mathbf{r};\Gamma)\big[{\rm e}^{\Sigma_t(\Gamma)}-1\big]\rangle_{{\rm leq},\boldsymbol{\lambda}_t}$.  We have a similar decomposition for the mean rate densities of the displacement field, leading to $\partial_t u^a({\bf r},t)+J^{\rm R}_{u^a}({\bf r},t)+J^{\rm D}_{u^a}({\bf r},t)=0$.

An  expansion in the gradients of the macrofields gives the dissipative current densities as ${J}^{{\rm D},a}_{c^\alpha}  = {\mathcal{L}}^{ab}_{\alpha\beta}{\nabla}^b{\lambda}^\beta+{\cal O}(\nabla^2)$. The linear response coefficients $ {\mathcal{L}}^{ab}$ are obtained as the following Green-Kubo formulas,
\begin{align}
\label{eq:GK}
{\mathcal{L}}^{ab}_{\alpha\beta} =\frac{1}{V} 
\int_0^{\infty}{\rm d}t \, \langle \delta \hat{\mathbb J}^{\prime a}_{c^\alpha}(t)\, \delta \hat{ \mathbb J}^{\prime b}_{c^\beta}(0)\rangle_{\rm eq}\;,
\end{align}
where the global currents $\delta \hat{ \mathbb J}^{\prime a}_{c^\alpha}(t) \equiv \int {\rm d}{\bf r} \, \delta \hat{J}^{\prime a}_{c^\alpha}({\bf r},t)$ are given by $\delta \hat{J}^{\prime a}_{c^\alpha} \equiv \delta \hat{J}^a_{c^\alpha}-\widehat{P}_{\boldsymbol{\lambda}_{\rm eq}}\delta \hat{J}^a_{c^\alpha}$ in terms of the deviations of the current densities with respect to their equilibrium mean values, $\delta \hat{J}^a_{c^\alpha} \equiv \hat{J}^a_{c^\alpha}- \langle \hat{J}^a_{c^\alpha} \rangle_{\rm eq}$, and some projector $\widehat{P}_{\boldsymbol{\lambda}_{\rm eq}}$ onto the subspace of the slow variables in the limit where the conjugate fields take their equilibrium values \cite{MG20,MG21,MG23a}. The symmetric part of the matrix of the linear response coefficients ${\cal L}^{ab,{\rm S}}_{\alpha \beta}=({\cal L}^{ab}_{\alpha \beta}+{\cal L}^{ba}_{ \beta\alpha})/2$ can be expressed in terms of Einstein-Helfand formulas as
\begin{align}
\label{eq:HE}
{\mathcal{L}}^{ab,{\rm S}}_{\alpha\beta} =\lim_{t\to\infty} \frac{1}{2tV} 
\langle  \hat{\mathbb G}^{\prime a}_{c^\alpha}(t)\,  \hat{ \mathbb G}^{\prime b}_{c^\beta}(t)\rangle_{\rm eq}\;,
\end{align}
where $\hat{\mathbb G}^{\prime a}_{c^\alpha}(t)\equiv \int_0^t {\rm d}\tau \, \delta \hat{ \mathbb J}^{\prime a}_{c^\alpha}(\tau)$ are the so-called Helfand moments \cite{H60}. The entropy production rate can thus be evaluated with the quadratic form $k_{\rm B}^{-1}{{\rm d}S}/{{\rm d}t}=\int {\rm d}{\bf r} \, {{\cal L}}^{ab,{\rm S}}_{\alpha\beta}\, {\nabla}^a{\lambda}^\alpha\, {\nabla}^b{\lambda}^\beta$, which is non-negative in agreement with the second law of thermodynamics.

For one-component crystals, the Green-Kubo formulas of all the possible transport coefficients have been listed in reference~\cite{MG21}.  They include the viscosity coefficients $\eta^{abcd}$, the heat conductivities $\kappa^{ab}$, the coefficients of vacancy diffusion $\zeta^{ab}$ and vacancy thermodiffusion $\xi^{ab}$, and coefficients describing possible coupling between the transport of momentum to those of heat $\chi^{abc}$ or vacancies $\theta^{abc}$.  

\subsection{Perfect crystals}
\label{sec:pftcrystal}

In order to obtain the hydrodynamic equations for one-component crystals, we have first to evaluate the reversible parts of the current densities.

A special role is played by the local conservation equation for mass, because the mass current density is identically equal to the momentum density:  $\hat{J}_\rho^a=\hat{g}^a$.  Since the macroscopic velocity field is defined as the velocity of the mass center in every element of the medium according to $v^a\equiv \langle\hat{g}^a\rangle_t/\langle\hat{\rho}\rangle_t$, the statistical average of the mass local conservation equation, i.e., $\partial_t\,  \langle\hat{\rho}\rangle_t  + {\nabla}^a \langle\hat{J}^a_{\rho}\rangle_t  = 0$, exactly gives the continuity equation $\partial_t\, \rho  + {\nabla}^a (\rho\, v^a) = 0$ with $\rho\equiv\langle\hat{\rho}\rangle_t$.  Another consequence is that the mean value of the mass current density over the exact probability distribution is equal to its mean value with respect to the local-equilibrium one, so that $\langle\hat{J}^a_{\rho}\rangle_t= J^{{\rm R},a}_{\rho}$ and the dissipative part of the mass current density is necessarily equal to zero, $J^{{\rm D},a}_{\rho}=0$.  Moreover, we note that the mean value of the energy density can also be evaluated, giving $\langle\hat{\epsilon}\rangle_t\equiv\epsilon=\epsilon_0+\rho{\bf v}^2/2$, where $\epsilon_0$ is the internal energy density in the frame moving with the element of the medium at the velocity ${\bf v}=(v^a)$.

The reversible parts of the energy and momentum current densities have been evaluated in references~\cite{MG20,MG21}, showing that
\bea
&& J^{{\rm R},a}_{\epsilon} \equiv \langle\hat{J}^a_{\epsilon}\rangle_{{\rm leq},\boldsymbol{\lambda}_t} = \epsilon \, v^a - \sigma^{ab}\, v^b + O(\nabla^2) \, , \\
&& J^{{\rm R},a}_{g^b} \equiv \langle\hat{J}^a_{g^b}\rangle_{{\rm leq},\boldsymbol{\lambda}_t} = \rho \, v^a \, v^b - \sigma^{ab} + O(\nabla^2) \, , 
\eea
where $\sigma^{ab}=-p\, \delta^{ab}+\phi^{ab}$ is the reversible stress tensor, $p$ denoting the hydrostatic pressure.  Furthermore, the reversible part of the rate densities for the displacement field have been shown to be given by
\be
J^{\rm R}_{u^a} \equiv \langle\hat{J}_{u^a}\rangle_{{\rm leq},\boldsymbol{\lambda}_t} = -v^a
\ee
in appendix A of reference~\cite{MG21}.

Here, we introduce the assumption of the {\it perfect crystal} as a crystal where the dissipative part of the rate densities for the displacement field is equal to zero: $J^{\rm D}_{u^a}=0$.  This assumption can be justified in crystals where vacancy diffusion and vacancy thermodiffusion are negligible because much smaller than the diffusivities associated with viscosity and heat conduction.  Vacancies are vacant lattice sites due to the movement of atoms from the lattice sites towards interstitial sites and ultimately towards the surface of the crystal.  In the presence of vacancies, the mean mass density in a strained crystal can be expressed on scales larger than the lattice primitive unit cells as $\rho=\rho_{{\rm eq},0}-\rho_{{\rm eq},0}\nabla^a u^a-mc$, where $\rho_{{\rm eq},0} = \frac{m}{v} \int_v {\rm d}{\bf r} \, n_{\rm eq}({\bf r})$ and $c$ is the vacancy concentration.  For a perfect crystal, where $\partial_t u^a = v^a$, we find that the vacancy concentration remains invariant in time since $m\partial_t c=-\partial_t \rho-\rho_{{\rm eq},0}\, \partial_t \nabla^a u^a=0$, as a consequence of the linearized continuity equation $\partial_t\rho\simeq - \rho_{{\rm eq},0}\nabla^a v^a$.  In particular, in the hard-sphere crystal, the vacancies are known to have diffusivities significantly smaller than the diffusivities of viscosity and heat conduction \cite{BA71,KDEP90,DvBK21} and, moreover, to be rare with a number fraction of the order of $10^{-4}$ or below \cite{KK04,L20}.  Accordingly, the transport coefficients of vacancy diffusion, vacancy thermodiffusion, and of coupling between vacancy and momentum transports can be taken to be negligible ($\zeta^{ab}\simeq 0$, $\xi^{ab}\simeq 0$, and $\theta^{abc}\simeq 0$), so that the approximation $J^{\rm D}_{u^a}\simeq 0$ is justified.

Furthermore, the hard-sphere crystal is a fcc crystal, which is a centrosymmetric cubic crystal of crystallographic point group ${\rm O}_{\rm h}$, where all third-rank tensors are equal to zero, so that there is no possible coupling between the transports of vectorial and tensorial quantities in this crystal ($\theta^{abc}=0$ and $\chi^{abc}=0$).  Therefore, the only transport properties that may exist in a perfect fcc crystal are the viscosities $\eta^{abcd}$ and the heat conductivities $\kappa^{ab}$.  Because of microreversibility, they satisfy the Onsager reciprocal relations, $\eta^{abcd}=\eta^{cdab}$ and $\kappa^{ab}=\kappa^{ba}$.  In this case, the transport coefficients form a symmetric matrix and, thus, the Green-Kubo formulas~\eqref{eq:GK} coincide with the Einstein-Helfand formulas~\eqref{eq:HE}: ${\mathcal{L}}^{ab}_{\alpha\beta}={\mathcal{L}}^{ab,{\rm S}}_{\alpha\beta}$.

Therefore, the linearized set of hydrodynamic equations for the perfect hard-sphere crystal reads
 \begin{align}
\partial_t  \delta \rho & = - \rho \, \nabla^a  \delta v^a  \, , \label{macro-eq-rho}\\
 \partial_t \delta \epsilon_0  & = -(\epsilon_0+p) \, \nabla^a \delta v^a + \kappa^{ab} \, \nabla^a\nabla^b \delta T \, , \label{macro-eq-e0}\\
\rho\, \partial_t \delta v^b  & = \nabla^a  \delta\sigma^{ab} + \eta^{abcd}\, \nabla^a\nabla^c \delta v^d  \, , \label{macro-eq-v}\\
\partial_t  \delta u^{a} & = \delta v^{a} \, , \label{macro-eq-u}
\end{align}
ruling the time evolution of the deviations of mass density $\delta\rho$, internal energy density $\delta\epsilon_0$, velocity $\delta v^b$, and displacement $\delta u^a$ with respect to the equilibrium rest state of the crystal, where the mass density is equal to $\rho=\rho_{{\rm eq},0}$, the internal energy density $\epsilon_0$, and the hydrostatic pressure $p$.  In order to make predictions, we now need to compute the equations of state for the internal energy $\epsilon_0$ and the pressure $p$, the elastic properties given by the stress-strain relation between the stress tensor $\delta\sigma^{ab}$ and the strain tensor $\delta u^{ab}$, and the values of the transport coefficients $\eta^{abcd}$ and $\kappa^{ab}$, on the basis of the microscopic dynamics.

\subsection{Helfand moments for perfect crystals in the $(N,V,E)$-ensemble}
\label{sec:HScrystal}

In one-component perfect crystals, the transport properties to consider are the momentum and energy transports characterized by their respective current densities, global currents, and Helfand moments.  Here, our goal is to calculate the expressions of the Helfand moments for momentum and energy transports in the $(N,V,E)$-ensemble used in molecular dynamics simulations.

\paragraph{Transport of momentum.} The Helfand moments associated with momentum transport  are given by $\hat{\mathbb G}^{\prime ab}(t) \equiv \int_0^t {\rm d} \tau \, \delta {\mathbb J}^{\prime ab} (\tau)$ in terms of the global momentum currents $\delta {\mathbb J}^{\prime ab}\equiv\int  {\rm d}{\bf r}\, \delta \hat{J}^{\prime a}_{g^b}({\bf r};\Gamma)$.  As aforementioned, these global currents are  defined by the volume integrals of the microscopic momentum current densities after projecting out the slow variables, $\delta \hat{J}^{\prime a}_{g^b} \equiv \delta\hat{J}_{g^b}^a - \widehat{P}_{\boldsymbol{\lambda}_{\rm eq}}\delta\hat{J}_{g^b}^a$.  As shown in references~\cite{MG20,MG21,MG23a}, the latter can be expressed as
\begin{align}
\delta \hat{J}^{\prime a}_{g^b} & = \delta\hat{J}_{g^b}^a +\left(\frac{\partial \sigma^{ab}}{\partial \epsilon_0}\right)_{\rho, \boldsymbol{\mathsf u}}  \delta \hat{\epsilon} + \left(\frac{\partial \sigma^{ab}}{\partial \rho}\right)_{\epsilon_0, \boldsymbol{\mathsf u}} \delta \hat{\rho} + \left(\frac{\partial \sigma^{ab}}{\partial u^{cd}}\right)_{\epsilon_0, \rho} \delta \hat u^{cd}\, ,
\label{eq:mom_crnt_dens_prime}
\end{align}
where the bare microscopic momentum current densities read
\begin{align}
\hat{J}^a_{g^b}({\bf r};\Gamma) &=\sum_i \frac{ p_i^{a}\,  p_i^{b}}{m}  \, \delta({\bf r}-{\bf r}_{i}) + \frac{1}{2} \sum_{i\neq j} r^a_{ij} F_{ij}^{b} \int_0^1 {\rm d}\xi \, \delta\left(\mathbf{r} - \mathbf{r}_i + \xi \, \mathbf{r}_{ij}\right)
\label{eq:mom_crnt_dens}
\end{align}
with ${\bf r}_{ij}\equiv{\bf r}_{i}-{\bf r}_{j}$.  We note that the equilibrium mean value of equation~\eqref{eq:mom_crnt_dens} gives the stress tensor of the crystal at equilibrium according to $\langle \hat{J}^a_{g^b}\rangle_{\rm eq} = -\sigma^{ab}$.  To obtain the global currents $\delta {\mathbb J}^{\prime ab}$, we use that
\begin{align}
& \int  {\rm d}{\bf r}\, \hat{J}^{a}_{g^b} ({\bf r};\Gamma) = \sum_i \frac{ p_i^{a}\,  p_i^{b}}{m} + \frac{1}{2} \sum_{i\neq j} r^a_{ij} F_{ij}^{b} \;,
& \int  {\rm d}{\bf r}\, & \langle \hat{J}^{a}_{g^b}({\bf r};\Gamma)\rangle_{\rm eq} = -\sigma^{ab}V\,,  \\
& \int  {\rm d}{\bf r}\, \delta \hat{\epsilon} ({\bf r};\Gamma)= \sum_i \varepsilon_i - E = E-E =0\;,
& \int  {\rm d}{\bf r}\, & \delta \hat{\rho} ({\bf r};\Gamma) = \sum_i 1 - N = N-N =0\;, 
\end{align}
where the last two equations are consequences of the conservation of the total energy $E$ and total particle number $N$ in the $(N,V,E)$-ensemble we consider.
For the  term involving the deviation $\delta \hat{u}^{ab}$ of the linear strain tensor, we first use equation~\eqref{eq:displu} to get
\begin{align}
\hat{u}^{ab} &= -\frac{{\rm i}}{2} \int {\rm d}{\bf r'}   \int_{\cal B} \frac{{\rm d}{\bf k}}{(2\pi)^3} \left[k^a (\pmb{\cal N}^{-1})^{bc} +k^b (\pmb{\cal N}^{-1})^{ac} \right]{\rm e}^{{\rm i} {\bf k}\cdot({\bf r}-{\bf r'})} \, {\nabla'}^c n_{\rm eq}({\bf r'}) \left[\hat n({\bf r'};\Gamma)-n_{\rm eq}({\bf r'})\right] .
\end{align}
The equilibrium mean value of this quantity vanishes $\left\langle \hat{u}^{ab}\right\rangle_{\rm eq} =0$, since $\langle\hat n({\bf r'};\Gamma)-n_{\rm eq}({\bf r'})\rangle_{\rm eq}= n_{\rm eq}({\bf r'})-n_{\rm eq}({\bf r'})= 0$, leading to $\delta\hat{u}^{ab}=\hat{u}^{ab}$. Next, we have that $\int  {\rm d}{\bf r}\, \delta\hat{u}^{ab}  ({\bf r};\Gamma)=0$, because $\int  {\rm d}{\bf r}\, {\rm e}^{{\rm i} {\bf k}\cdot({\bf r}-{\bf r'})} = (2\pi)^3 \delta({\bf k})$ and $k^a\delta({\bf k})=0$. Therefore, the contributions of the extra terms in addition to $\delta\hat{J}_{g^b}^a$ in equation~\eqref{eq:mom_crnt_dens_prime} are all equal to zero, i.e., the projector on the slow variables does not contribute to the global currents, $\int {\rm d}{\bf r}\, \widehat{P}_{\boldsymbol{\lambda}_{\rm eq}}\delta\hat{J}_{g^b}^a=0$.  Finally, we find that $\delta {\mathbb J}^{\prime ab}\equiv\int  {\rm d}{\bf r}\, \delta \hat{J}^{\prime a}_{g^b}=\int  {\rm d}{\bf r}\, \delta \hat{J}^{a}_{g^b}\equiv\delta {\mathbb J}^{ab}$ in the $(N,V,E)$-ensemble. The Helfand moments associated with the transport of momentum  are thus given by 
\begin{align}
{\mathbb G}^{ab}(t)&= \int_0^t{\rm d}\tau\left(\sum_i\frac{p_i^ap_i^b}{m}+ \frac{1}{2}\sum_{i \ne j} r_{ij}^a {F}^b_{ij}\right) .\label{eq:Gab}
\end{align}
Since there is no contribution from the projector, the prime has been removed.  We note that the equilibrium mean values of the Helfand moments \eqref{eq:Gab} give the reversible stress tensor according to $\left\langle {\mathbb G}^{ab}(t) \right\rangle_{\rm eq} =-\sigma^{ab}Vt$.

\paragraph{Transport of heat.} The Helfand moments associated with energy transport are given by ${\mathbb G}_{\epsilon}^{\prime a}(t) \equiv \int_0^t {\rm d}\tau \, \delta {\mathbb J}^{\prime a}_{\epsilon} (\tau)$ in terms of the global energy current $\delta {\mathbb J}^{\prime a}_{\epsilon}\equiv\int  {\rm d}{\bf r}\, \delta \hat{J}^{\prime a}_{\epsilon}({\bf r};\Gamma)$. The corresponding microscopic current density has the following expression~\cite{MG20,MG21},
\begin{align}
\delta \hat{J}^{\prime a}_{\epsilon} & = \delta \hat{J}^a_{\epsilon} -\rho^{-1}(\epsilon_0+p)\, \delta\hat{g}^a\, ,
\end{align}
where 
\begin{align}
\hat{{J}}_{\epsilon}^{a}(\mathbf{r};\Gamma) \equiv \sum_i \frac{p_i^a}{m}\, \varepsilon_i \, \delta(\mathbf{r} - \mathbf{r}_i)  + \frac{1}{2}\sum_{i\neq j} {r}_{ij}^a {F}^b_{ij} \frac{{p}^b_i + {p}^b_j}{2m}\, \int_{0}^1 {\rm d}\xi\, \delta\left(\mathbf{r} - \mathbf{r}_i + \xi\, \mathbf{r}_{ij}\right) .
\end{align}
Since $\int  {\rm d}{\bf r}\, \delta \hat{g}^a ({\bf r};\Gamma) = \sum_i {\bf p}_i - {\bf P} = {\bf P}-{\bf P} =0$, we directly obtain the Helfand moments associated with the transport of energy as
\begin{align}
{\mathbb G}_{\epsilon}^a(t)&=\int_0^t {\rm d}\tau \left( \sum_{i}  \frac{{p}^a_i}{m}\, \varepsilon_{i} + \frac{1}{2} \sum_{i\neq j}r^a_{ij}{F}^{b}_{ij} \frac{{p}^b_{i}+{p}^b_{j}}{2m} \right) .\label{eq:Geq}
\end{align}
We note that the equilibrium mean value of the Helfand moments \eqref{eq:Geq} are equal to zero, $\left\langle {\mathbb G}_{\epsilon}^a(t) \right\rangle_{\rm eq} =0$,  since they are odd in the momenta of the particles.

\subsection{Equilibrium properties of the perfect crystal}
\label{sec:equil}

The equilibrium properties of the crystal include the equations of state for the internal energy and the hydrostatic pressure together with the thermodynamic quantities that can be derived therefrom, as well as the properties of elasticity~\cite{W98}.

\paragraph{Internal energy.} In the frame moving with the medium element, the mean internal energy density is given by $\epsilon_0=\langle\hat\epsilon\rangle_{\rm eq}$ and the mean internal energy per unit mass by $e_0=\epsilon_0/\rho$.
The specific heat capacity at constant volume is deduced from the equation of state for the internal energy by $c_v\equiv(\partial e_0/\partial T)_v$.

\paragraph{Hydrostatic pressure.} The hydrostatic pressure is obtained from the equilibrium statistical averages of the Helfand moments associated with the transport of momentum as
\begin{align}
p&=\lim_{t\rightarrow\infty}\frac{1}{3Vt}\left\langle{\mathbb G}^{aa}(t)\right\rangle_{\rm eq}\,. \label{eq:pressure}
\end{align}
Furthermore, if the mean values of the Helfand moments are carried out in the cubic domain $[0,L[^3$, we should expect to obtain the diagonal matrix $\langle {\mathbb G}^{ab}(t) \rangle_{\rm eq} =pVt \delta^{ab}$, because second-order tensors are diagonal in the cubic geometry.

Having obtained the hydrostatic pressure, we can compute the specific heat capacity at constant pressure $c_p\equiv(\partial e_0/\partial T)_p$, the isothermal bulk modulus
\begin{align}
B_T&\equiv -V\left(\frac{\partial p}{\partial V}\right)_T= n\left(\frac{\partial p}{\partial n}\right)_T ,
\label{eq:B_T}
\end{align}
and the adiabatic bulk modulus $B_S$ given by a similar formula but deriving at constant entropy rather than at constant temperature.  The bulk moduli are the inverses of the corresponding compressibilities.  Moreover, these quantities are related to each other by the specific heat ratio
\begin{align}
\gamma&\equiv\frac{c_p}{c_v}=\frac{B_S}{B_T}\; .
\label{eq:c_p}
\end{align}

\paragraph{Elastic constants.}
The isothermal elastic constants are computed by considering the stress-strain relation after a deformation of the simulation domain~\cite{W98}. Let ${\bf R}$ be the reference configuration of the crystal, for instance, as simulated with periodic boundary conditions on the cubic domain $[0,L[^3$. As aforementioned, the stress tensor is diagonal giving the hydrostatic pressure in this configuration, $\sigma^{ab}({\bf R},T)=-p\, \delta^{ab}$.  Let ${\bf r}={\bf R}+{\bf u}({\bf R})$ be a new configuration deformed by the linear transformation $r^a=D^{ab}R^b$, where $\boldsymbol{\mathsf D}$ is the deformation matrix.  For this deformation, the corresponding homogeneous nonlinear Lagrangian strain tensor is defined as
\begin{align}
u^{ab}& \equiv \frac{1}{2}\left( \boldsymbol{\mathsf D}^{\rm T}\cdot\boldsymbol{\mathsf D} -\boldsymbol{\mathsf 1}\right)^{ab} \, .
\label{eq:nonlin_Lagrangian_strain}
\end{align}
Since small deformations are here considered, the nonlinear strain tensor is well approximated by the linear one of the local-equilibrium approach.  In this new configuration, the stress tensor can be computed according to
\begin{align}
\sigma^{ab}({\bf r},T)&=- \lim_{t\rightarrow\infty}\frac{1}{Vt}\left\langle{\mathbb G}^{ab}(t)\right\rangle_{\rm eq}\label{eq:stress_tensor}
\end{align}
using the periodic boundary conditions~\eqref{eq:deformed_pbc} of the deformed spatial domain.  The so-obtained stress tensor is not expected to be diagonal since the simulation domain is no longer cubic.  Under the effects of small enough deformations, we expect a linear relationship between the non-diagonal deviations of the stress tensor and the corresponding strain tensor.  The elastic constants can thus be computed with the stress-strain relation
\begin{align}
\sigma^{ab}({\bf r},T)&=\sigma^{ab}({\bf R},T)+ B_T^{abcd}\, u^{cd} \, ,
\end{align}
where $ B_T^{abcd}$ are the isothermal stress-strain coefficients~\cite{W98}. The latter are related to the isothermal elastic constants $C_T^{abcd}$ and the pressure $p$ as
\begin{align}
B_T^{abcd}&= C_T^{abcd}-p\left(\delta^{ac}\delta^{bd}+\delta^{ad}\delta^{bd}-\delta^{ab}\delta^{cd}\right) .
\end{align}
In cubic crystals, there are only three independent coefficients given by $B^T_{11}=C^T_{11}-p$, $B^T_{12}=C^T_{12}+p$, and $B^T_{44}=C^T_{44}-p$, as written in Voigt's notations presented in appendix~\ref{app:tensors}. The  isothermal bulk modulus can thus be obtained as~\cite{W98}
\begin{align}
B_T &=\frac{1}{3}\left(B^T_{11}+2B^T_{12}\right)=\frac{1}{3}\left(C^T_{11}+2C^T_{12}+p\right) .
\label{eq:Cbulkm}
\end{align}
Moreover, the adiabatic and isothermal stress-strain coefficients satisfy
\begin{align}
B_{11}^S-B_{11}^T=B_{12}^S-B_{12}^T=B_S-B_T 
\qquad\mbox{and}\qquad
B_{44}^S-B_{44}^T=0 \, .
\end{align}
Since $B_S=\gamma\, B_T$, the adiabatic elastic constants are related to the isothermal constants according to
\begin{align}
C_{11}^S&=C_{11}^T+(\gamma-1) B_T \;,&&C_{12}^S=C_{12}^T+(\gamma-1)B_T\;,&&C_{44}^S=C_{44}^T \, .
\end{align}

The elastic constants are computed numerically from the stress tensors of some undeformed and deformed configurations, using the statistical averages of the Helfand moments~\eqref{eq:Gab}. For $C^T_{11}$ and $C^T_{12}$, we take a stretch for the deformation matrix and their relation~\eqref{eq:Cbulkm} to the isothermal bulk modulus.  For $C^T_{44}$, we perform a shear deformation. More details on the evaluation of the elastic constants in a cubic crystal are given in appendix~\ref{app:EC}.

\subsection{Nonequilibrium properties of the perfect crystal}
\label{sec:nonequil}

As explained in subsection~\ref{sec:pftcrystal}, the nonequilibrium properties to consider in perfect crystals are the viscosities and the heat conductivities.

The viscosity tensor is given by the covariances of the Helfand moments~\eqref{eq:Gab} as
\begin{align}
\eta^{abcd}&=\lim_{t\rightarrow\infty}\frac{1}{2t k_{\rm B}TV}\left\langle\left[ {\mathbb G}^{ab}(t)- \left\langle{\mathbb G}^{ab}(t)\right\rangle_{\rm eq}\right]\left[{\mathbb G}^{cd}(t)- \left\langle{\mathbb G}^{cd}(t)\right\rangle_{\rm eq}\right]\right\rangle_{\rm eq}\, .\label{eq:etaH}
\end{align}
In cubic crystals, this fourth-rank tensor can be expressed in terms of only three coefficients, which are denoted $\eta_{11}$, $\eta_{12}$, and $\eta_{44}$ in Voigt's notations, as discussed in appendix~\ref{app:tensors}.

The heat conductivity tensor is calculated with the covariances of the Helfand moments~\eqref{eq:Gab} as
\begin{align}
\kappa^{ab} &=\lim_{t\rightarrow\infty}\frac{1}{2t k_{\rm B}T^2V} \left\langle {\mathbb G}_{\epsilon}^a(t)\,{\mathbb G}_{\epsilon}^b(t)\right\rangle_{\rm eq}\,.
	\label{eq:kappaH}
\end{align}
In cubic crystals, second-rank tensors are diagonal, so that a single heat conductivity is defined therein, $\kappa^{ab}=\kappa\, \delta^{ab}$.

\section{Results for the perfect hard-sphere crystal}
\label{sec:Results_HS}

\subsection{The hard-sphere dynamics}
\label{sec:HSmodel}
The hard-sphere model consists of a dynamical system of $N$ hard spheres of diameter $d$ and mass $m$ moving in a cubic domain of volume $V=L^3$ with periodic boundary conditions.  In the Hamiltonian function~\eqref{Hamilt_fn}, the binary energy potential ${\cal V}({r}_{ij})$ is equal to zero for interparticle distances such that ${r}_{ij}>d$ and is otherwise infinite.  If the particles do not overlap, i.e., if $r_{ij} >d$ for all the particles $1\leq i \neq j\leq N$, then the Hamiltonian function reduces to the total kinetic energy.  The total energy and the total linear momentum are conserved by the hard-sphere dynamics and read
\begin{align}\label{eq:HS-csts-motion}
H=\sum_{i=1}^N \frac{{\bf p}_i^2}{2m}
\qquad\mbox{and}\qquad
{\bf P} = \sum_{i=1}^N {\bf p}_i \, .
\end{align}

The particles move in free flights interrupted by instantaneous elastic collisions occurring at successive times $\{t_c\}$ between pairs of hard spheres. The free flights of the particles can be written as
\begin{align}\label{eq:free_flight}
{\bf r}_i(t) = {\bf r}_i(t_c) + \frac{{\bf p}_i(t_c+0)}{m} \, (t-t_c) 
\qquad\mbox{for}\qquad
t_c < t< t_{c+1} \, .
\end{align} Upon elastic collisions, the positions remain continuous, ${\bf r}_{i}(t_c+0)={\bf r}_{i}(t_c-0)={\bf r}_{i}(t_c)\equiv {\bf r}_{i}^{(c)}$, but the momenta undergo the following discontinuous changes,
\begin{align}\label{eq:elastic_coll}
\left\{
\begin{array}{l}
{\bf p}_i(t_c+0) = {\bf p}_i(t_c-0) +\Delta{\bf p}_{ij}^{(c)} \, , \\ 
{\bf p}_j(t_c+0) = {\bf p}_j(t_c-0) -\Delta{\bf p}_{ij}^{(c)}\, , \\
\end{array}
\right.
\end{align}
where $\Delta{\bf p}_{ij}^{(c)}=- \frac{2}{d^2} \big[ {\bf r}_{ij}^{(c)}\cdot {\bf p}_{ij}(t_c-0)\big] {\bf r}_{ij}^{(c)}$ are the momentum exchanges, which are expressed in terms of the canonically conjugate positions ${\bf r}_{ij}={\bf r}_i-{\bf r}_j$ and momenta ${\bf p}_{ij}\equiv \frac{1}{2}\left({\bf p}_i-{\bf p}_j\right)$ of the two colliding particles.  We note that every collision $c$ involves a specific pair $i\ne j$ of particles and conserves the total energy and the total momentum of the two colliding particles.  The relative position ${\bf r}_{ij}^{(c)}$ is always taken as the one satisfying the minimum image convention~\cite{H97}.

The hard-sphere dynamics can be simulated using an event-driven algorithm \cite{H97}, in which the time evolution is driven by the successive elastic collisions undergone by the particles.  At every collision between two particles $i$, all the times for their possible future collisions with the other particles $k$ are computed by solving the quadratic equations $[{\bf r}_i(t)-{\bf r}_k(t)]^2=d^2$ for $t-t_c>0$.  For each sphere, the smallest time is selected for its next collision to happen.  

According to equation~(\ref{eq:elastic_coll}), the momenta of the colliding particles undergo the jumps $\Delta{\bf p}_{ij}^{(c)}=-\Delta{\bf p}_{ji}^{(c)}$, so that the force exerted on the $i^{\rm th}$ particle by the collisional dynamics can be written as ${\bf F}_{ij}(t) = \sum_{c} \Delta{\bf p}_{ij}^{(c)} \, \delta(t-t_c)$, where the sum extends over all the elastic collisions $\{c\}$.  As a consequence, the Helfand moments in equations~\eqref{eq:Gab} and \eqref{eq:Geq} become
\begin{align}
{\mathbb G}^{ab}(t)&= \sum_{(c\to c+1)\,\in\,[0,t]} \sum_i\frac{p_i^a(t_c+0)}{m}\, p_i^b(t_c+0) \, (t_{c+1}-t_c) + \sum_{c\, \in \, [0,t]}r_{ij}^{a (c)} \, \Delta {p}_{ij}^{b (c)} \,,\label{eq:GabHS}\\
{\mathbb G}_{\epsilon}^a(t)&=\sum_{(c\to c+1)\,\in\,[0,t]} \sum_{i} \frac{{p}^a_i(t_c+0) }{m}\, \frac{[{\bf p}_{i}(t_c+0)]^2}{2m}\, (t_{c+1}-t_c) + \sum_{c\, \in \, [0,t]}  r^{a (c)}_{ij} \, \Delta{p}_{ij}^{b (c)} \, \frac{{p}^{b (c)}_{i}+{p}^{b (c)}_{j}}{2m}\, ,\label{eq:GeaHS}
\end{align}
where the first sum extends to the full or partial free flights during the time interval $[0,t]$,  the second sum is carried out over the collisions happening during this time interval, and ${\bf p}_i^{(c)}+{\bf p}_j^{(c)}\equiv {\bf p}_i(t_c\pm 0)+{\bf p}_j(t_c\pm 0)$ because of momentum conservation in binary collisions. The Helfand moments associated with the transport of momentum  have the symmetry  ${\mathbb G}^{ab} = {\mathbb G}^{ba}$.

The total energy and the total linear momentum are fixed at the values $H=E$ and ${\bf P}=0$ given by the initial conditions.  At equilibrium, the temperature is thus related to the total energy by $k_{\rm B}T = (2/3)(E/N)$.  In the simulations, the hard spheres are taken of unit diameter $d=1$ and unit mass $m=1$, and the temperature has the value $k_{\rm B}T=1$. The particles are initially located in a fcc lattice. The numerical results are presented in terms of dimensionless quantities.  The dimensionless particle density is defined as $n_*\equiv nd^3=Nd^3/V$.  The dimensionless positions, momenta, and time are respectively given by ${\bf r}_{i*}\equiv {\bf r}_i/d$, ${\bf p}_{i*}\equiv{\bf p}_i/\sqrt{mk_{\rm B}T}$, and $t_*\equiv (t/d)\sqrt{k_{\rm B}T/m}$.   Using these rescaled quantities, the Helfand moments, the hydrostatic pressure, the stress tensor, and the transport coefficients can be expressed as
\bea
&&{\mathbb G}^{ab} = d\sqrt{mk_{\rm B}T} \, {\mathbb G}^{ab}_* \, , \quad {\mathbb G}_{\epsilon}^a = dk_{\rm B}T \, {\mathbb G}_{\epsilon *}^a \, , \nonumber\\
&& p = \frac{k_{\rm B}T}{d^3} \, p_* \, , \quad \sigma^{ab} = \frac{k_{\rm B}T}{d^3} \, \sigma^{ab}_* \, , \nonumber\\
&& \eta^{abcd} = \frac{m}{d^2}\sqrt{\frac{k_{\rm B}T}{m}} \, \eta^{abcd}_* \, , \quad \kappa^{ab} = \frac{k_{\rm B}}{d^2}\sqrt{\frac{k_{\rm B}T}{m}} \, \kappa^{ab}_*\;,
\eea
in terms of the corresponding dimensionless quantities.

After the initial conditions are fixed and before computing statistical averages and covariances, the dynamics is run during the transient time $t_{\rm transient}$ in order to reach the thermalization expected in equilibrium.  The parameters of the simulation are the transient time $t_{*\,{\rm transient}}=50$, the time step $\Delta t_*=0.01$, the number of time steps $n_{\rm steps}=100$, and the number of trajectories used for statistics $N_{\rm stat}=10^4$. We consider the densities $n_*\in\{1.037,1.1,1.2,1.3,1.4\}$, where $n_*=1.037$ is close to the melting density, and the following numbers of particles $N\in\{32,108,256,500\}$. More details on the simulation and the processing of data are provided in appendix~\ref{app:nummet}.

An important quantity for the hard-sphere dynamics is the collision frequency $\nu$, i.e., the mean number of collisions undergone by a particle per unit time.  As seen in the right panel of figure~\ref{Fig:Crystal}, the trajectories of the hard spheres are confined in small domains of size $\Delta\ell$, area $\Delta A\sim\Delta\ell^2$, and volume $\Delta V\sim\Delta\ell^3$.  The volume available for the motion of the particle shrinks as $\Delta V\sim\Delta\ell^3\sim d^3(\sqrt{2}-n_*)$ in the vicinity of the close-packing density $n_*=\sqrt{2}$.  Moreover, the time of flight between the collisions is of the order of $\Delta t\sim \Delta\ell/\bar{v}$, where $\bar{v}\sim\sqrt{k_{\rm B}T/m}$ is the mean thermal velocity.  Since the free flights are interrupted by collisions on the walls of the confining domains having area $\Delta A\sim\Delta\ell^2$, the collision frequency per particle should scale as $\nu\sim d^2/(\Delta A\Delta t)\sim d^2 \bar{v}/\Delta\ell^3\sim (\sqrt{2}-n_*)^{-1}$, which is the divergence observed in the numerical simulations near the close-packing density $n_*=\sqrt{2}$.  The values of the collision frequency per particle are given in table~\ref{Tab:collfreq} and its divergence is shown in the left panel of figure~\ref{Fig:divergences}.

\subsection{Equilibrium properties of the perfect hard-sphere crystal}
\label{sec:equprop}

For the hard-sphere crystal, the maximal particle density is reached for close-packing configurations of the spheres, where the density has the value $n_{*{\rm max}}=\sqrt{2}$.  Moreover, the coexistence zone between the fluid and crystalline phases extends over the density range, $0.938\pm 0.003 \leq n_* \leq 1.037\pm 0.003$, where the pressure has the dimensionless value $p_{*{\rm FC}}=11.55 \pm 0.11$ \cite{S98,footnote}. Therefore, hard-sphere crystals exist for the densities $1.037\pm 0.003 \leq n_* < \sqrt{2}$.  

\paragraph{Internal energy.} Since the mean kinetic energy only contributes to the internal energy in the hard-sphere dynamics,  the internal energies per unit volume and per unit mass are respectively given by $\epsilon_0=(3/2)nk_{\rm B}T$ and $e_0=\epsilon_0/\rho=3k_{\rm B}T/(2m)$, where $T$ is the temperature.  Therefore, the specific heat capacity at constant volume takes the value $c_v\equiv(\partial e_0/\partial T)_v=3k_{\rm B}/(2m)$.

\paragraph{Hydrostatic pressure and related properties.} In the hard-sphere system, the equation of state for the pressure can be factorized into the thermal energy and a function depending only on the particle density as
\begin{align}\label{eq:EoS_pressureHS}
p(n,T) = k_{\rm B} T \, f(n) 
\qquad\mbox{with} \qquad
\left(\frac{\partial f}{\partial T}\right)_n=0 \, .
\end{align}
The hydrostatic pressure is computed with equation~(\ref{eq:pressure}) in a cubic domain $[0,L[^3$, where $\sigma^{ab}=-p\, \delta^{ab}$.  Results are given in table~\ref{Tab:pressure} using the numerical methods of appendix~\ref{app:nummet}.

From the extrapolated values of the pressure  to $N\rightarrow \infty$, the equation of state for the pressure is obtained by fitting the data to the following analytical expression proposed by Speedy~\cite{S98},
\begin{align}
\label{eq:Ansatz_eos}
f_{\rm P}(n) = n\left(\frac{3}{1-z} -a \, \frac{z-b}{z-c}\right)
\qquad\mbox{with}\qquad
z\equiv n d^3 /\sqrt{2} \, .
\end{align}
In reference~\cite{S98}, the parameters are computed as $a = 0.5921$, $b = 0.7072$, and $c=0.601$. Figure~\ref{Fig:TH} shows that the extrapolated values  to $N\rightarrow \infty$ are in excellent agreement with Speedy's equation of state~\eqref{eq:Ansatz_eos}.  Fitting the analytical form~\eqref{eq:Ansatz_eos} to our numerical data gives the values $a = 0.5904, b = 0.7080$, and $ c = 0.603$, which confirms the excellent agreement. 

For the hard-sphere crystal, we infer from the hydrostatic pressure~\eqref{eq:EoS_pressureHS} and the specific heat capacity at constant volume, $c_v=3k_{\rm B}/(2m)$, that the isothermal bulk modulus, the specific heat capacity at constant pressure, and the specific heat ratio are respectively given by
\begin{align}
B_T&=n k_{\rm B} T f'(n)\, ,\label{eq:B_T}\\
c_p&=\frac{3 k_{\rm B}}{2m} \left[1+\frac{2}{3}\frac{f(n)^2}{n^2f'(n)}\right] , \label{eq:c_P}\\
\gamma&=1+\frac{2}{3}\frac{f(n)^2}{n^2f'(n)}\, , \label{eq:gamma}
\end{align}
where $f'(n)=({\mathrm d}/{\mathrm d}n)f(n)$. The numerical values of these quantities are given versus the particle density in table~\ref{Tab:TH} and shown in figure~\ref{Fig:TH} in comparison with the predictions of Speedy's equation of state.

The hydrostatic pressure and the isothermal bulk modulus steeply increase with the particle density near the close-packing density, as seen in figure~\ref{Fig:TH}.  The numerical data shows that the hydrostatic pressure diverges as $p\sim (\sqrt{2}-n_*)^{-1}$, as observed in the left panel of figure~\ref{Fig:divergences}. This divergence can be explained by considering the definition of the hydrostatic pressure as a force per unit area $p=\Delta F/\Delta A$ and the fact that the force is the momentum  transferred per unit time, $\Delta F=\Delta P/\Delta t$ by the collisions on the walls of the domains confining the trajectories plotted in the right panel of figure~\ref{Fig:Crystal}.  If the confining domains have the size $\Delta\ell$, their walls have the area $\Delta A\sim\Delta\ell^2$, and the time interval between the collisions goes as $\Delta t\sim\Delta\ell/\bar{v}$ with the mean thermal velocity $\bar{v}\sim\sqrt{k_{\rm B}T/m}$.  Since each collision transfers a momentum of the order of $\Delta P\sim m\bar{v}$, the hydrostatic pressure should behave as $p=\Delta F/\Delta A=\Delta P/(\Delta A\Delta t)\sim m\bar{v}^2/\Delta\ell^3 \sim (\sqrt{2}-n_*)^{-1}$, confirming the divergence that is observed in the numerical data.  As a consequence of equation~\eqref{eq:B_T}, the isothermal bulk modulus diverges as $B_T\sim (\sqrt{2}-n_*)^{-2}$, as confirmed in the left panel of figure~\ref{Fig:divergences}.  Moreover, equations~\eqref{eq:c_P} and~\eqref{eq:gamma} imply that the specific heat capacity at constant pressure and the specific heat ratio remain constant, because $c_p\sim\gamma\sim (\sqrt{2}-n_*)^{0}$ near the close-packing density $n_*=\sqrt{2}$, as also observed in the numerical results.

\paragraph{Elastic constants.}  They are calculated with the method described in subsection~\ref{sec:equil} and in appendix~\ref{app:EC}.  The stress tensor is computed using the statistical averages of the Helfand moments~\eqref{eq:GabHS} between the undeformed and some deformed configurations.  The details of the deformations used to obtain $C^T_{11}$, $C^T_{12}$, and $C^T_{44}$ are given in appendix~\ref{app:EC}, whereas the numerical methods and the choice of parameters are discussed in appendix~\ref {app:acovH}.
The numerical results are presented in table~\ref{Tab:EC}, showing a good agreement with the literature~\cite{FL87,PF03}. In figure~\ref{Fig:EC}, the extrapolated values to $N\rightarrow\infty$ of the elastic constants are shown as function of the particle density, along with a rational function
\begin{align}
{\cal R}(n)&=\frac{a_0+a_1n+a_2n^2}{1+b_1n}\label{eq:ratvisc}
\end{align}
fitted to the data.  The numerical results plotted in the left panel of figure~\ref{Fig:divergences} show that the elastic constants diverge as $C_{11}^T, C_{12}^T, C_{44}^{T}\sim (\sqrt{2}-n_*)^{-2}$ in the vicinity of the close-packing density $n_*=\sqrt{2}$, which is consistent with the fact that the elastic constants and the bulk moduli behave similarly since all of them are first-order response properties of the stress tensor with respect to strain.

We note that, for the hard-sphere system, we have that $B_S-B_T=(\gamma-1)B_T={2p^2}/{(3n k_{\rm B}T)}$, whereupon the adiabatic elastic constants are related to the isothermal constants according to~\cite{W98}
\begin{align}
C_{11}^S&=C_{11}^T+\frac{2p^2}{3n k_{\rm B} T}\;,&&C_{12}^S=C_{12}^T+\frac{2p^2}{3n k_{\rm B}T}\;,&&C_{44}^S=C_{44}^T\, .
\end{align}

\subsection{Transport properties of the perfect hard-sphere crystal}
\label{sec:transport_PHS}
The viscosity tensor given by equation~\eqref{eq:etaH} is computed with the Helfand moments of equation~\eqref{eq:GabHS}. Since the hard-sphere crystal is cubic, there are three independent coefficients, which can be evaluated with
\begin{align}
\eta_{11} &=\frac{1}{3}\left(\eta^{xxxx}+\eta^{yyyy}+\eta^{zzzz}\right) ,&&
\eta_{12} =\frac{1}{3}\left(\eta^{xxyy}+\eta^{xxzz}+\eta^{yyzz}\right) ,&&
\eta_{44} = \frac{1}{3}\left(\eta^{yzyz}+\eta^{zxzx}+\eta^{xyxy}\right) .\label{eq:eta44}
\end{align}
As explained in appendix~\ref{app:acovH}, we have verified that the viscosity tensor has indeed the expected cubic symmetry of equation~\eqref{eq:cub4tens}. The numerical results for the viscosities versus the total particle number $N$ are given in table~\ref{Tab:eta} using the numerical methods of appendix~\ref{app:nummet}. The extrapolated values to $N\rightarrow\infty$ are shown as a function of the density in figure~\ref{Fig:TC}, along with a rational function~\eqref{eq:ratvisc} fitted to the data. 
 
The heat conductivity tensor given by equation~\eqref{eq:kappaH} is computed with the Helfand moments of equation~\eqref{eq:GeaHS}.  As discussed in appendix~\ref{app:acovH}, we  first verify that the tensor is diagonal, as expected for a cubic symmetry. The heat conductivity is thus evaluated with
\begin{align}
\kappa&=(\kappa^{xx}+\kappa^{yy}+\kappa^{zz})/3\;.
\label{eq:kappa}
\end{align}
The numerical results for the heat conductivity versus $N$ are given in table~\ref{Tab:kappa} using the numerical methods of appendix~\ref{app:nummet}.  The extrapolated values to $N\rightarrow\infty$ are shown as a function of the density in figure~\ref{Fig:TC}. The results are consistent with earlier works~\cite{GAW70,PBHB20}.  Moreover, in the last column of table~\ref{Tab:kappa} and in figure~\ref{Fig:TC}, the numerical results are compared with the values for the heat conductivity predicted with Enskog's theory (see appendix~\ref{app:Enskog}), showing good agreement.

In cubic crystals, heat conduction looks isotropic as in fluids because there is a single coefficient for the second-rank tensor of heat conductivity.  However, since the viscosity tensor has rank four, there are three viscosities in cubic crystals, but only two in isotropic fluids.  If the phase was isotropic, the three viscosities of the cubic crystal would be given by $\eta_{11}=\zeta+\frac{4}{3}\eta$, $\eta_{12}=\zeta-\frac{2}{3}\eta$, and $\eta_{44}=\eta$ in terms of the shear $\eta$ and bulk $\zeta$ viscosities of the fluid phase.  However, the isotropy condition $\eta_{11}-\eta_{12}=2\eta_{44}$ is not satisfied by the numerical data of table~\ref{Tab:eta}, confirming the anisotropy of the viscosity tensor in the cubic crystal.  Comparing the viscosities $\zeta+\frac{4}{3}\eta$ and $\eta$ of the fluid phase~\cite{MG23b} to the corresponding viscosities $\eta_{11}$ and $\eta_{44}$ of the crystalline phase, we note that they undergo a drop across the fluid-crystal coexistence zone from the end $n_*=0.938$ of the fluid phase to the beginning $n_*=1.037$ of the crystalline phase.  In this regard, the crystalline order tends to decrease the values of the transport coefficients across the phase transition.

 Furthermore, in figure~\ref{Fig:TC}, we observe that the viscosities $\eta_{11}$, $\vert\eta_{12}\vert$, and $\eta_{44}$, as well as the heat conductivity $\kappa$ rapidly increase with the density in the vicinity of close packing, reaching much higher values than in the fluid phase \cite{MG23b}.  The numerical data show that the transport coefficients diverge as $\eta_{11}, \eta_{12}, \eta_{44}, \kappa \sim (\sqrt{2}-n_*)^{-1}$ near the maximal close-packing density $n_*=\sqrt{2}$, as seen in the right panel of figure~\ref{Fig:divergences}.  The divergence of the heat conductivity as $\kappa \sim (\sqrt{2}-n_*)^{-1}$ can be explained on the basis of Enskog's kinetic theory summarized in appendix~\ref{app:Enskog}, as shown in reference~\cite{PBHB20}.  In general, these divergences find their origin in the proportionality of the transport coefficients to the collision frequency, which diverges similarly as $\nu\sim (\sqrt{2}-n_*)^{-1}$  for the hard-sphere crystal.

\section{Conclusion and perspectives}
\label{sec:conclusion}

We have obtained the hydrodynamic properties of the perfect hard-sphere crystal from the microscopic dynamics using the statistical-mechanical method of  Helfand moments.  These properties are the ones needed to establish the macroscopic equations of motion for the hydrodynamics of crystals.  They include the equilibrium or reversible properties such as the hydrostatic pressure, the isothermal bulk modulus, the specific heat capacities, and the isothermal elastic constants, which are given in terms of the mean kinetic energy and the mean values of the Helfand moments associated with momentum transport.  The hydrodynamic properties also include the nonequilibrium or irreversible transport properties.  Since  one-component perfect crystals are empty of vacancies and vacancy diffusion is absent therein, the transport properties to consider are the viscosities and the heat conductivities, which are given by Einstein-Helfand formulas in terms of the covariances of the Helfand moments associated with the transports of momentum and energy.  Because the hard-sphere crystal is face-centered cubic, there are three elastic constants, three viscosities, and one heat conductivity.  We have computed all these properties  as a function of the particle density in the crystalline phase extending from the melting density $n_*\simeq 1.037$ up to the maximal close-packing density $n_*=\sqrt{2}$.

For the equilibrium properties, our results are in good to excellent agreement with the values previously published in the literature for the hydrostatic pressure \cite{S98} and for the elastic constants \cite{FL87,PF03}.  For the heat conductivity, our results also agree with earlier works~\cite{GAW70,PBHB20}.  In this paper, we have furthermore computed the values of the viscosities, which play an essential role in the damping of the sound modes in the crystal. To our knowledge, the viscosities of crystals have not been computed before by Green-Kubo or Einstein-Helfand formulas.

We have investigated in detail the cubic symmetries of the tensors of heat conductivities and viscosities.  The computations based on the Helfand moments clearly show that the second-rank tensor of heat conductivities is diagonal in the cubic hard-sphere crystal.  Furthermore, only 12 among the 36 components of the fourth-rank tensor of viscosities are nonzero and these 12 nonzero components are equal to either one or another of the three viscosities $\eta_{11}$, $\eta_{12}$, and $\eta_{44}$ in Voigt's notations, as expected for cubic crystals.

As compared to the values of the hydrostatic pressure, the isothermal bulk modulus, and the transport coefficients in the fluid phase~\cite{MG23b}, we observe that they undergo a drop across the fluid-crystal coexistence interval of densities $0.938\leq n_* \leq 1.037$, but they increase with the density in the crystalline phase for the densities $1.037 \leq n_* \leq\sqrt{2}$.  Furthermore, these properties steeply increase and diverge near the close-packing density $n_*=\sqrt{2}$.  Near this maximal possible density of the hard-sphere crystal, the hydrostatic pressure and the collision frequency per particle diverge as $(\sqrt{2}-n_*)^{-1}$.  Consistently, the isothermal bulk modulus and the elastic constants diverge as $(\sqrt{2}-n_*)^{-2}$.  Moreover, we show that the transport coefficients diverge as $(\sqrt{2}-n_*)^{-1}$ for the reason that they typically scale as the collision frequency.

In the following two papers~\cite{MG23c,MG23d}, all these properties will be used to solve the hydrodynamic equations of the perfect hard-sphere crystal in order to predict the analytic forms of the correlation and spectral functions characterizing the dynamics of density and momentum fluctuations around equilibrium, as well as the poles of the spectral functions at the complex frequencies of the hydrodynamic modes of the crystal.  These predictions will be compared to the numerical results of molecular dynamics simulations of the perfect hard-sphere crystal with the purpose of testing the local-equilibrium approach in the crystalline phase, as we did for the fluid phase in reference~\cite{MG23b}.


\section*{Acknowledgements}

The authors thank James Lutsko for helpful discussions about computational methods used for crystals.  We acknowledge the support of the Universit\'e Libre de Bruxelles (ULB) and the Fonds de la Recherche Scientifique de Belgique (F.R.S.-FNRS) in this research. J.~M. is a Postdoctoral Researcher of the Fonds de la Recherche Scientifique de Belgique (F.R.S.-FNRS).  Computational resources have been provided by the Consortium des Equipements de Calcul Intensif (CECI), funded by the Fonds de la Recherche Scientifique de Belgique (F.R.S.-FNRS) under Grant No. 2.5020.11 and by the Walloon Region.


\appendix

\section{Fourth-rank tensors and Voigt’s notation in cubic symmetry}
\label{app:tensors}

A fourth-rank tensor with the symmetries $T^{abcd}=T^{bacd}=T^{abdc}$ has only 36 components, which can be written in the following matrix form:
\begin{align}
\boldsymbol{\mathsf T}&\equiv
	 \left(\begin{array}{cccccc} 
	 T^{xxxx} &  T^{xxyy} &  T^{xxzz} & T^{ xxyz} &  T^{xxzx} & T^{xxxy} \\
T^{yyxx} & T^{yyyy} &  T^{yyzz} &  T^{yyyz} &  T^{yyzx} &  T^{yyxy}\\
T^{zzxx} & T^{zzyy} &  T^{zzzz} &  T^{zzyz }&  T^{zzzx} &  T^{zzxy}\\
T^{yzxx} & T^{yzyy} & T^{yzzz} & T^{yzyz} & T^{yzzx }&  T^{yzxy}\\
T^{zxxx } & T^{zxyy } & T^{zxzz } & T^{zxyz } & T^{zxzx } & T^{zxxy}\\
T^{xyxx} & T^{xyyy} & T^{xyzz} & T^{xyyz} & T^{xyzx} &T^{xyxy}
	 \end{array}\right) .
\end{align}
We may introduce the Voigt notations according to the substitutions:
\begin{align}
xx \to 1 \, , &&
yy \to 2 \, , &&
zz \to 3 \, , &&
yz \to 4 \, , &&
zx \to 5 \, , &&
xy \to 6 \, ,
\end{align}
and the tensor becomes
\begin{align}
\boldsymbol{\mathsf T}&=
\left(\begin{array}{cccccc} 
T_{11} &  T_{12} &  T_{13} &  T_{14} &  T_{15} & T_{16} \\
T_{21} & T_{22} & T_{ 23} &  T_{24} &  T_{25} &  T_{26}\\
T_{31} & T_{32} &  T_{33} &  T_{34} &  T_{35} &  T_{36}\\
T_{41} & T_{42} & T_{43} & T_{44} & T_{45} &  T_{46}\\
T_{51} & T_{52} & T_{53} & T_{54} & T_{55}& T_{56}\\
T_{61} & T_{62} & T_{63} & T_{64} & T_{65} &T_{66}
	 \end{array}\right) .
\end{align}
If there is a cubic symmetry, the tensor reduces to 
\begin{align}
\label{eq:cub4tens}
\boldsymbol{\mathsf T}&=
	 \left(\begin{array}{cccccc} 
	 T_{11} &  T_{12} &  T_{12} & 0&  0 &0 \\
T_{12} & T_{11} & T_{ 12} &0 & 0 & 0\\
T_{12} & T_{12} &  T_{11} & 0 & 0 & 0\\
0 & 0 & 0 & T_{44} & 0 &  0\\
0 & 0 & 0 & 0 & T_{44}& 0\\
0 & 0 & 0 & 0 & 0 &T_{44}
	 \end{array}\right) ,
\end{align}
where 
\begin{align}
	T_{11} &=\frac{1}{3}\left(T^{xxxx}+T^{yyyy}+T^{zzzz}\right) ,&&
	T_{12} =\frac{1}{3}\left(T^{xxyy}+T^{xxzz}+T^{yyzz}\right) ,&&
	T_{44} = \frac{1}{3}\left(T^{yzyz}+T^{zxzx}+T^{xyxy}\right) .
\end{align}

\section{Evaluation of the elastic constants in a cubic crystal}
\label{app:EC}


The initial undeformed configuration ${\bf R}$ has no strain,  so that $\boldsymbol{\mathsf D}=\boldsymbol{\mathsf 1}$, $u^{ab}=0$, and $\sigma^{ab}({\bf R},T)=-p\, \delta^{ab}$. The stress $\tilde\sigma^{ab}=\sigma^{ab}({\bf r},T)$ is then computed for a deformed configuration with a non-trivial $\boldsymbol{\mathsf D}$. 

To compute $C^T_{11}$ and $C^T_{12}$,  the stretch is given by the following deformation matrix, 
\begin{align}
\label{eq:dmatstretch}
\boldsymbol{\mathsf D}= \left(\begin{array}{cccccc} 
1+\delta &  0 &   0  \\
0 &   1-\delta &  0\\
0 & 0 &  \frac{1}{1-\delta^2}	 \end{array}\right) ,
\end{align}
where $\delta$ is an arbitrary parameter, small enough to be in the linear regime. This deformation changes the lengths of the edges of the simulation box from $L_x$ to $(1+\delta)L_x$, $L_y$ to $(1-\delta)L_y$, and $L_z$ to $L_z/(1-\delta^2)$. The nonlinear Lagrangian strain~\eqref{eq:nonlin_Lagrangian_strain} thus reads
\begin{align}
\boldsymbol{\mathsf u} = \left(\begin{array}{cccccc} 
\frac{1}{2}\left[\left(1+\delta\right)^2-1\right] &  0 &   0  \\
0 &   \frac{1}{2}\left[\left(1-\delta\right)^2-1\right]  &  0\\
0 & 0 &  \frac{1}{2}\left[\left(1-\delta^2\right)^{-2}-1\right] 	 \end{array}\right) .
\label{eq:strain-C11-C12}
\end{align}
The stress-strain relation gives
\begin{align}
C^T_{11}-C^T_{12}&=\frac{1}{3}\left(\frac{\tilde\sigma^{xx}-\tilde\sigma^{yy}}{u^{xx}-u^{yy}}+\frac{\tilde\sigma^{xx}-\tilde\sigma^{zz}}{u^{xx}-u^{zz}}+\frac{\tilde\sigma^{yy}-\tilde\sigma^{zz}}{u^{yy}-u^{zz}}\right)+2p\;.\label{eq:C11mC12}
\end{align}
We compute numerically $C^T_{11}-C^T_{12}$ from equation~\eqref{eq:C11mC12} and $C^T_{11}+2C^T_{12}$ with the bulk modulus using equation~\eqref{eq:Cbulkm} with the hydrostatic pressure for the hard-sphere crystal.

The computation of $C^T_{44}$ requires a strain matrix $\boldsymbol{\mathsf u}$ with at least one non-vanishing off-diagonal element.  The deformation matrix of a shear deformation is 
\begin{align}
\boldsymbol{\mathsf D}= \left(\begin{array}{cccccc} 
1 &  0 &   \delta  \\
0 &   1 &  0\\
0 & 0 &  1	 \end{array}\right) ,
&&
\text{such that } &&
\boldsymbol{\mathsf u} = \left(\begin{array}{cccccc} 
0 &  0 &   \delta/2  \\
0 &   0  &  0\\
\delta/2& 0 & \delta^2/2	 \end{array}\right) .
\label{eq:deform+strain-C44}
\end{align}
This choice of deformation matrix corresponds to a shift in the $x$-coordinates from $x$ to $x+z\,\delta$. The periodic boundary conditions and the measure of the distance in the $x$-direction must include the effects of the shear  geometry.  Here, the stress-strain relation gives
\begin{align}
C^T_{44}&=\frac{\tilde\sigma^{zx}}{2 \, u ^{zx}}+p \, .
\label{eq:C44-zx}
\end{align}
Similar formulas can be obtained for shear deformations in the two other directions, giving
\begin{align}
C^T_{44}=\frac{\tilde\sigma^{xy}}{2 \, u^{xy}}+p 
\qquad\mbox{and} \qquad
C^T_{44}=\frac{\tilde\sigma^{yz}}{2 \, u^{yz}}+p \, ,
\label{eq:C44-yz+zx}
\end{align}
if $u^{xy}\ne 0$ and $u^{yz}\ne 0$, respectively.

\section{Numerical methods}
\label{app:nummet}


\subsection{Averages and covariances of Helfand moments}
\label{app:acovH}

The event-driven algorithm computes the averages of the Helfand moments and  their covariances as function of time. The Helfand moments are calculated using equations~\eqref{eq:GabHS} and~\eqref{eq:GeaHS}. The covariances are computed with the Welford algorithm~\cite{W62}. The ensemble averages are obtained by taking the average over $N_{\text{stat}}$ trajectories of length $n_{\text{steps}}\Delta t$, where  $\Delta t$ is the time step at which the Helfand moments are computed and  $n_{\text{steps}}$ is the number of time steps in each trajectory. $N_{\text{stat}}$,  $n_{\text{steps}}$, $\Delta t$, and the transient time $t_{\text{transient}}$ are input parameters. The total elapsed time of the simulation is then $t_{\text{transient}}+N_{\text{stat}}n_{\text{steps}}\Delta t$.

\begin{figure}[h!]\centering
{\includegraphics[width=0.4\textwidth]{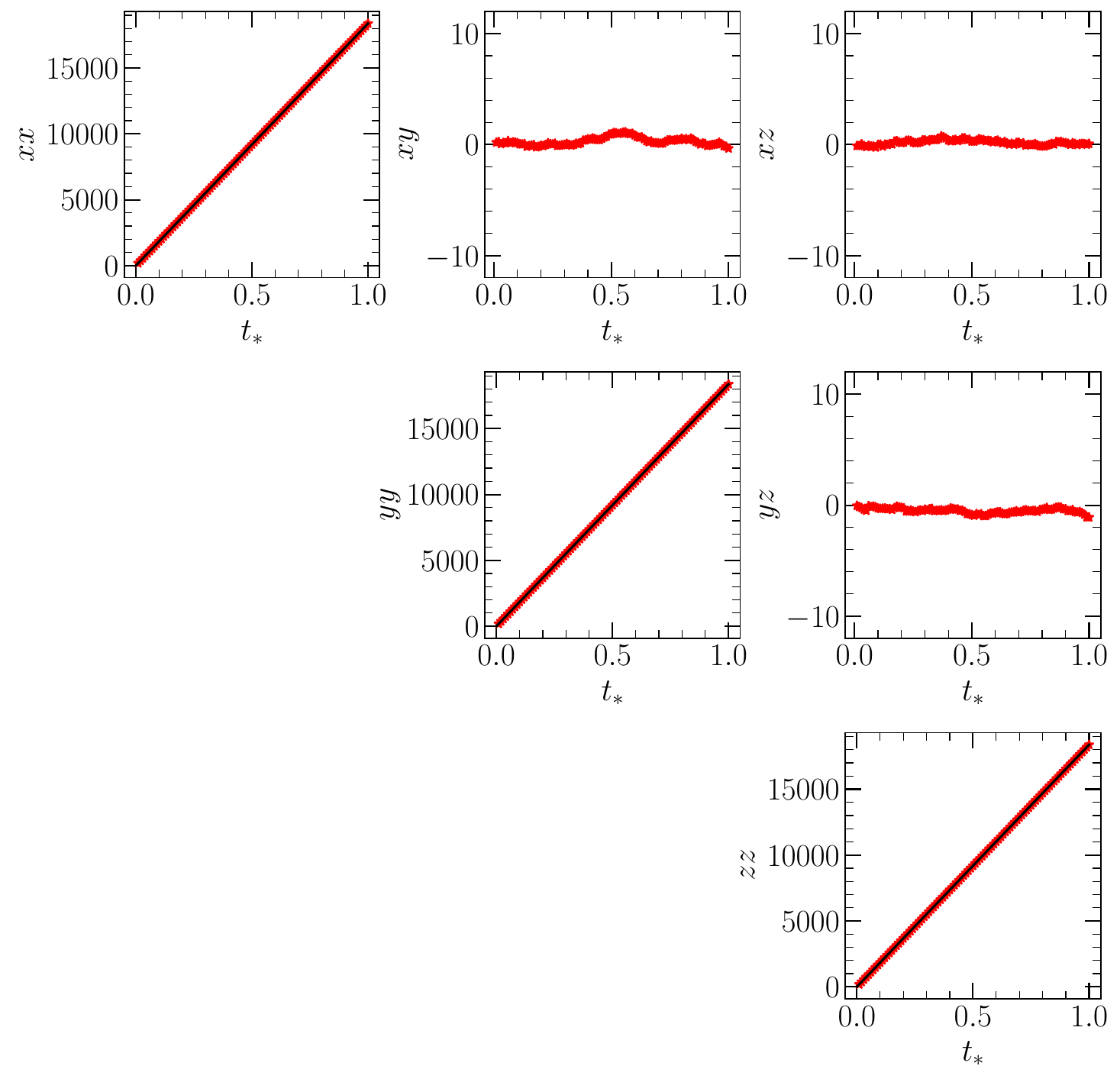}}
\caption[] {The statistical averages of the Helfand moments $\mathbb{G}^{ab}$ versus time from the molecular dynamics simulation (symbols). The solid black lines show a linear fit to the data.  The labels of the vertical axes denote the coordinates $ab$ of the Helfand moment $\mathbb{G}^{ab}$ that is shown in the plot.  The parameter values are $n_*=1.3$, $N=500$, $\Delta t_*=0.01$, $n_{\text{steps}}=100$, and $N_{\text{stat}}=10^4$.
}\label{Fig:AvgG}
\end{figure}

\begin{figure}[h!]\centering
{\includegraphics[width=0.5\textwidth]{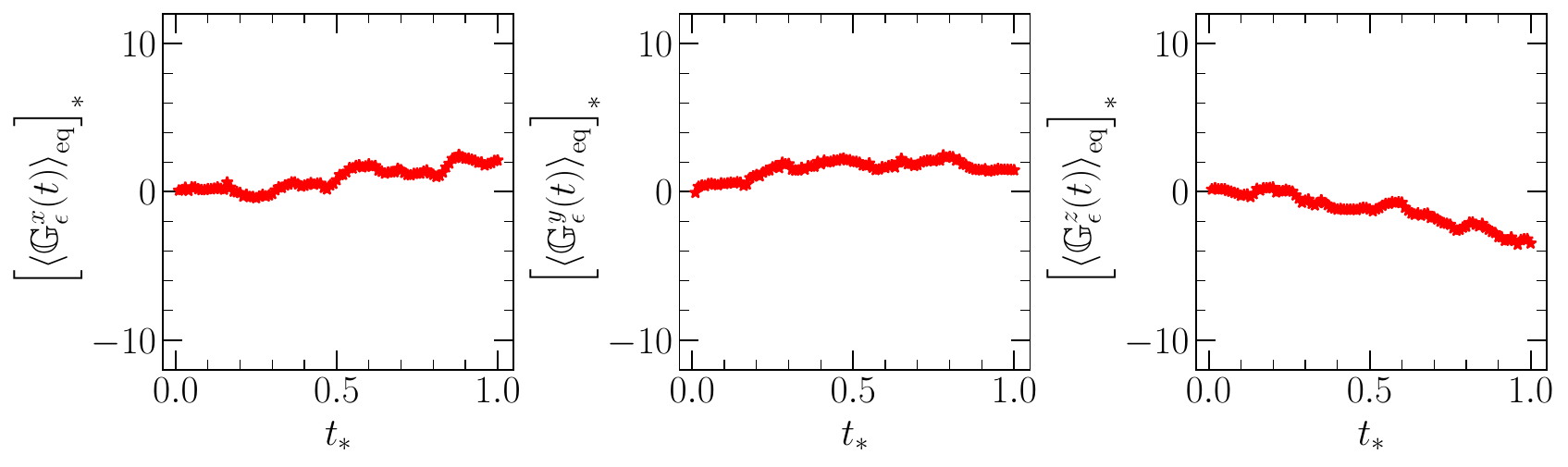}}
\caption[] {The statistical averages of the Helfand moments $\mathbb{G}_\epsilon^{a}$ versus time from the molecular dynamics simulation (symbols). The parameter values are $n_*=1.3$, $N=500$, $\Delta t_*=0.01$, $n_{\text{steps}}=100$, and $N_{\text{stat}}=10^4$.
}\label{Fig:AvgGe}
\end{figure}

From the average values of the Helfand moments, we first verify the cubic symmetry. In figure~\ref{Fig:AvgG}, the averages of the Helfand moments associated with momentum $\left\langle {\mathbb G}^{ab}(t)\right\rangle_{\rm eq}=-Vt\sigma^{ab}$ are shown to form a diagonal tensor, which is expected since the stress tensor is diagonal  having the form $\sigma^{ab}=-p\, \delta^{ab}$, when using simulations in the cubic domain $[0,L[^3$ for the computation of the hydrostatic pressure $p$. Moreover, the diagonal elements  show a linear dependence on time. In figure~\ref{Fig:AvgGe}, the averages of the Helfand moments associated with energy are shown to vanish,  $\left\langle {\mathbb G}_{\epsilon}^a(t)\right\rangle_{\rm eq}=0$, which is also expected at equilibrium since the Helfand moments~\eqref{eq:Geq} are odd in the momenta of the particles.

To extract the hydrostatic pressure, a  linear least squares fitting is performed with the time dependent entries $xx$, $yy$, and $zz$ of $\left\langle {\mathbb G}^{ab}(t)\right\rangle_{\rm eq}$. We obtain three values  for the pressure $\{p^{(i)}\pm\Delta p^{(i)}_{\text{LR}}\}$, where $i\in\{xx,yy,zz\}$ and $\Delta p^{(i)}_{\text{LR}}$ is the error made on the linear regression, which is estimated from the difference between values obtained by considering  the whole time interval and its second half. The hydrostatic pressure is obtained as $p=\sum_{i}p^{(i)}/3$ and the  error on $p$ is estimated as $\Delta p=\sqrt{\sum_{i}(\Delta p^{(i)})^2/6}$, where $\Delta p^{(i)}=|p^{(i)}-p|+|\Delta p^{(i)}_{\text{LR}}|$.

The  constants of elasticity are obtained from the averages of the Helfand moments $\left\langle {\mathbb G}^{ab}(t)\right\rangle_{\rm eq}$ under a strain given by the deformation matrix \eqref{eq:dmatstretch} for $C_{11}^T$ and $C_{12}^T$, or by the matrix \eqref{eq:deform+strain-C44} for $C_{44}^T$. The elastic constants are then calculated using the isothermal bulk modulus $B_T$ in equation~\eqref{eq:Cbulkm} together with equations~\eqref{eq:C11mC12} and \eqref{eq:C44-zx}. The simulation is run for several values of $\delta$, chosen as $\delta_*\in \{0.001,0.003,0.005,0.008,0.01\}$ for the stretch~\eqref{eq:dmatstretch} and $\delta_*\in \{0.001,0.012,0.0015,0.0018,0.002\}/L_*$ for the shear~\eqref{eq:deform+strain-C44}. Taking the average over the values obtained with different choices of $\delta$ gives the elastic constants reported in table~\ref{Tab:EC}. The errors are estimated as the standard deviations.

\begin{figure}[h!]\centering
{\includegraphics[width=.7\textwidth]{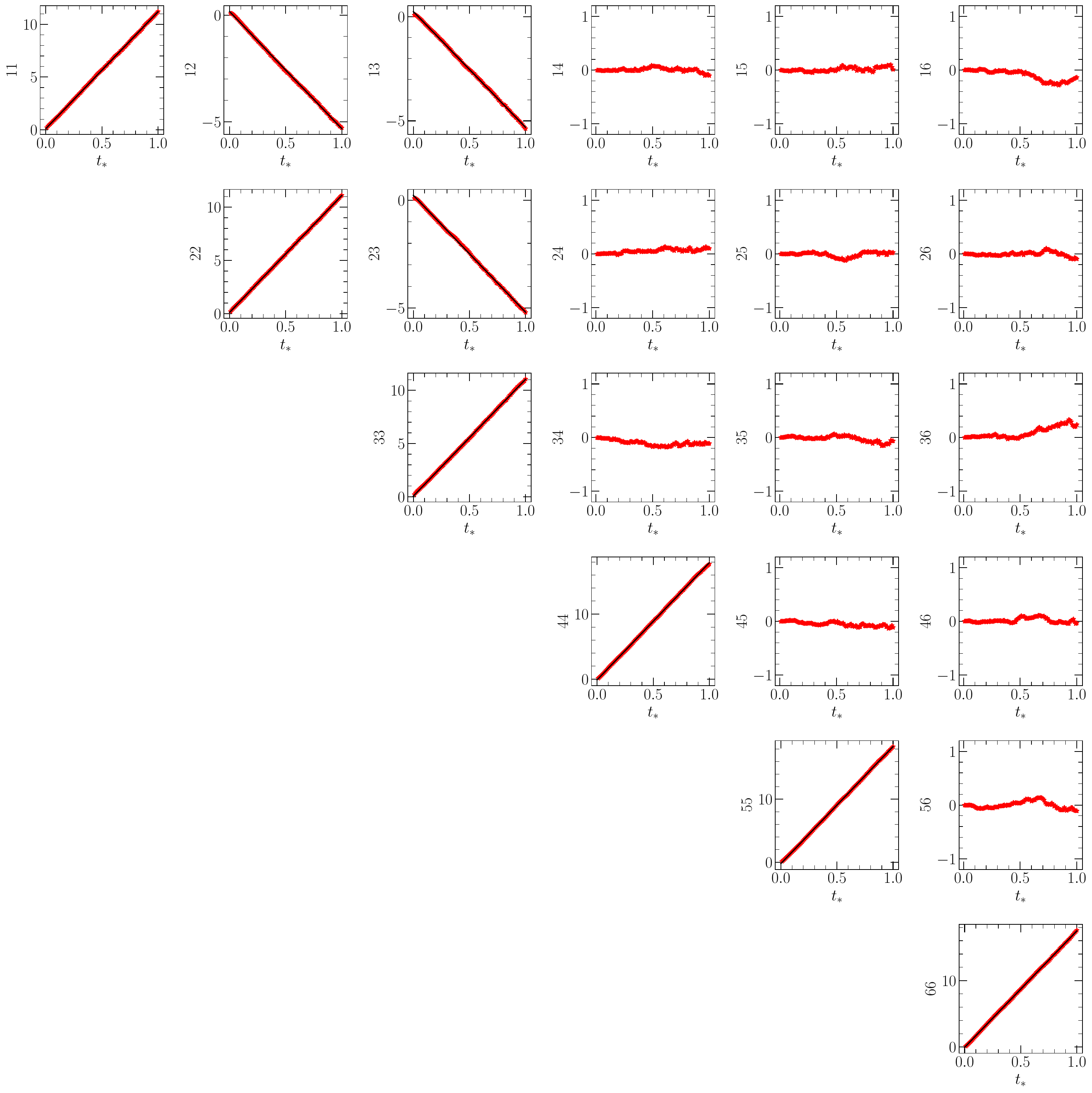}}
\caption[] {The covariances of the Helfand moments $\mathbb{G}^{ab}$ divided by $(2k_{\rm B}TV)$ versus time from the molecular dynamics simulation (symbols). The solid black lines show a linear fit to the data.  The label of the vertical axes are in Voigt's notation, $11$ corresponding to $xxxx$, $12$ to $xxyy$, ... for the coordinates $abcd$ of the covariance $\left\langle\left[ {\mathbb G}^{ab}(t)- \left\langle{\mathbb G}^{ab}(t)\right\rangle_{\rm eq}\right]\left[{\mathbb G}^{cd}(t)- \left\langle{\mathbb G}^{cd}(t)\right\rangle_{\rm eq}\right]\right\rangle_{\rm eq}$ shown in the plot. The parameter values are $n_*=1.3$, $N=500$, $\Delta t_*=0.01$, $n_{\text{steps}}=100$, and $N_{\text{stat}}=10^4$.

}\label{Fig:CovG}
\end{figure}

\begin{figure}[h!]\centering
{\includegraphics[width=0.35\textwidth]{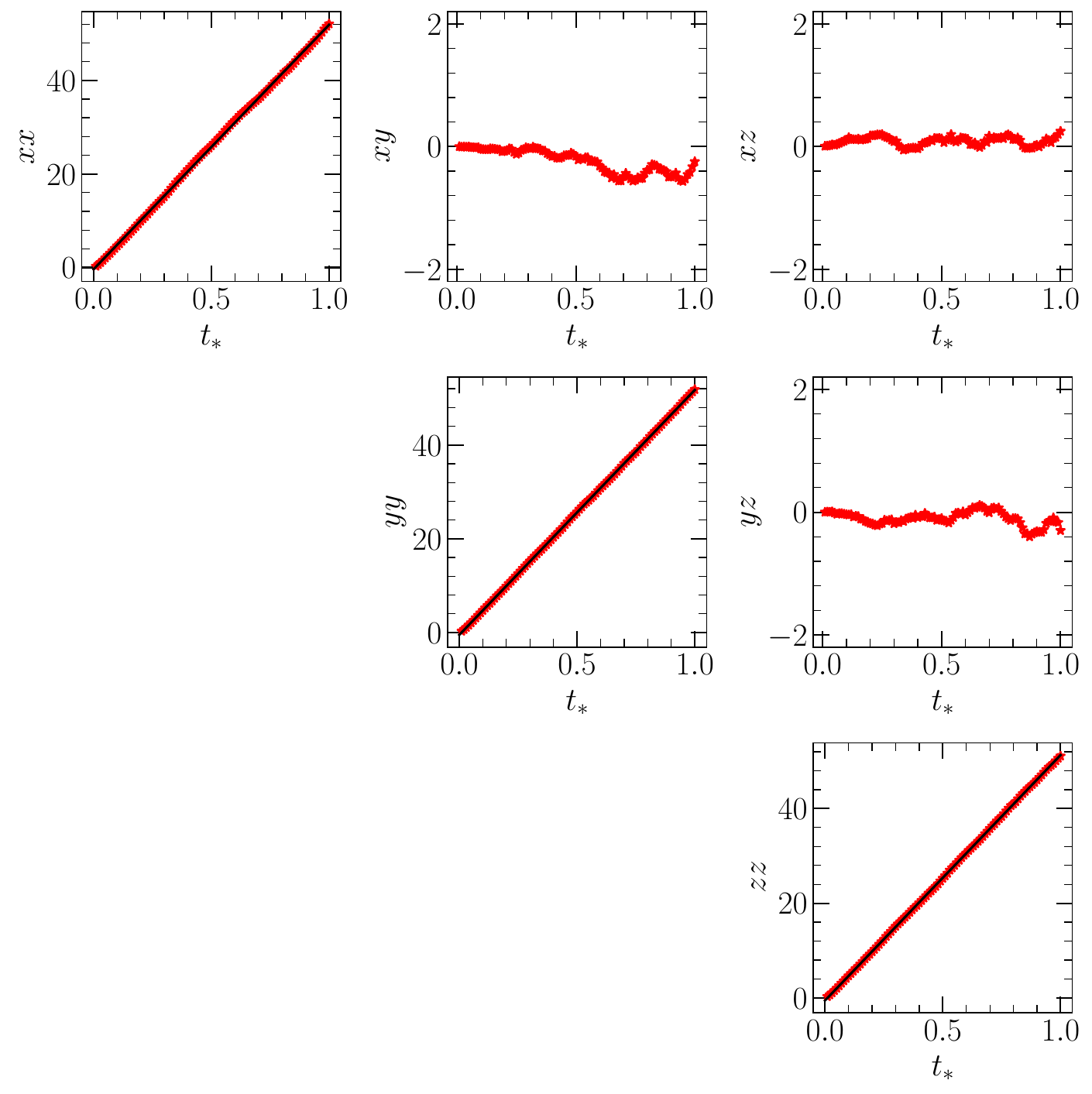}}
\caption[] {The covariances of the Helfand moments $\mathbb{G}^{a}_\epsilon$ divided by $(2k_{\rm B}T^2V)$ versus time from the molecular dynamics simulation (symbols). The solid black lines show a linear fit to the data. The labels of the vertical axes denote the coordinates $ab$ of the covariance $\left\langle {\mathbb G}_{\epsilon}^a(t)\,{\mathbb G}_{\epsilon}^b(t)\right\rangle_{\rm eq}$ shown in the plot.  The parameter values are $n_*=1.3$, $N=500$, $\Delta t_*=0.01$, $n_{\text{steps}}=100$, and $N_{\text{stat}}=10^4$.
}\label{Fig:CovGe}
\end{figure}

From the average values of the covariances of the Helfand moments, we also first verify the cubic symmetry. In figure~\ref{Fig:CovG} , the viscosity tensor~\eqref{eq:etaH} is shown to have the form of equation~\eqref{eq:cub4tens}. In figure~\ref{Fig:CovGe},  the heat conductivity tensor~\eqref{eq:kappaH} is shown to be diagonal.  The viscosities $\eta_{11}$, $\eta_{12}$, $\eta_{44}$, and the heat conductivity $\kappa$  are then extracted from the covariances of the Helfand moments, using equations~\eqref{eq:eta44} and~\eqref{eq:kappa} and the same method as for the hydrostatic pressure.


\subsection{Extrapolation to infinite $N$}
\label{app:infN}

\begin{figure}[h!]\centering
{\includegraphics[width=0.8\textwidth]{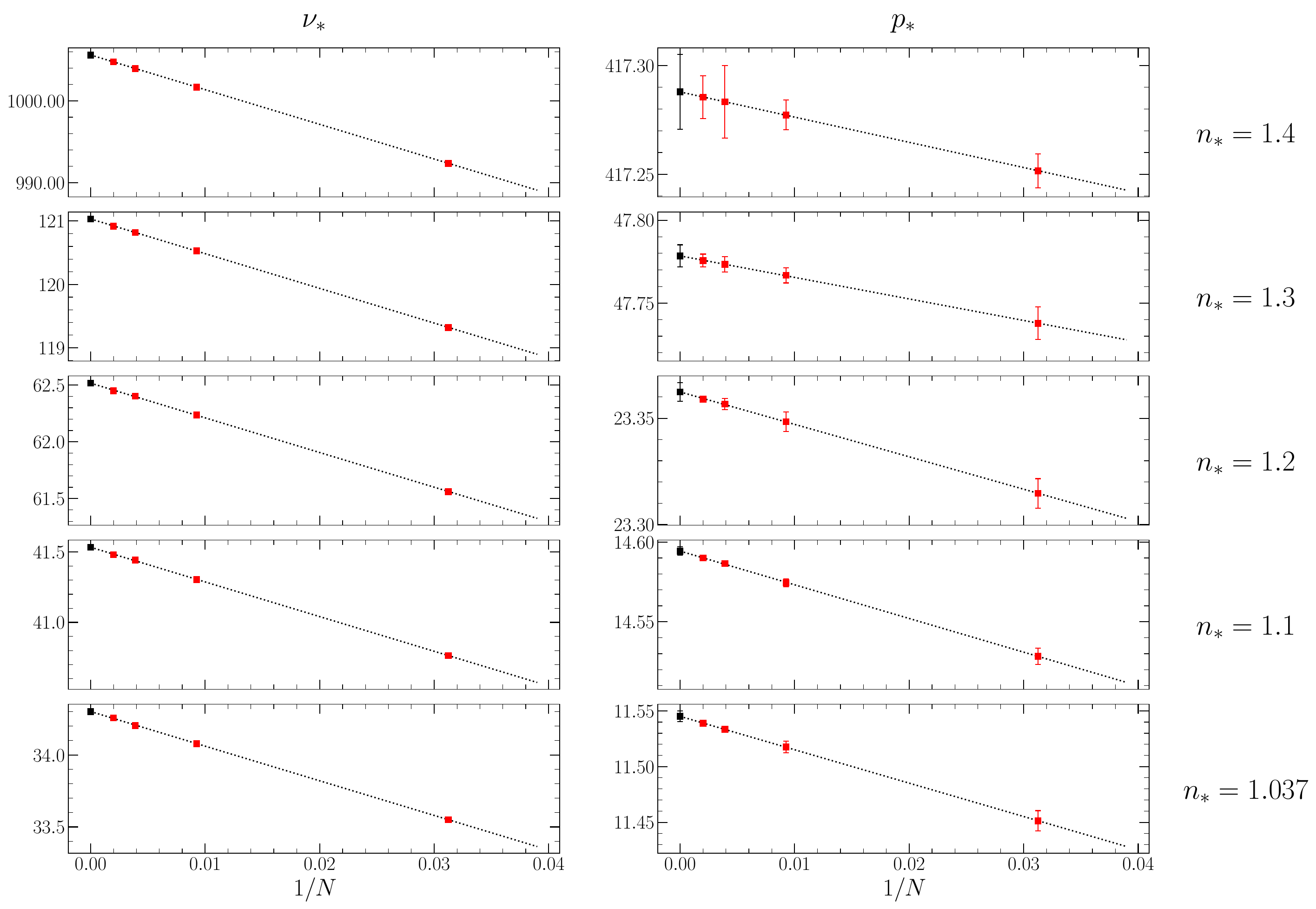}}
\caption[] {Collision frequency $\nu$ (table~\ref{Tab:collfreq}) and pressure $p$ (table~\ref{Tab:pressure}) versus the dependence $N^{-1}$ on the particle number $N$ and the density $n_*$ (symbols) with the linear regression approximation (dotted lines) for the extrapolation $N\to\infty$.}\label{Fig:pressure}
\end{figure}

\begin{figure}[h!]\centering
{\includegraphics[width=1.\textwidth]{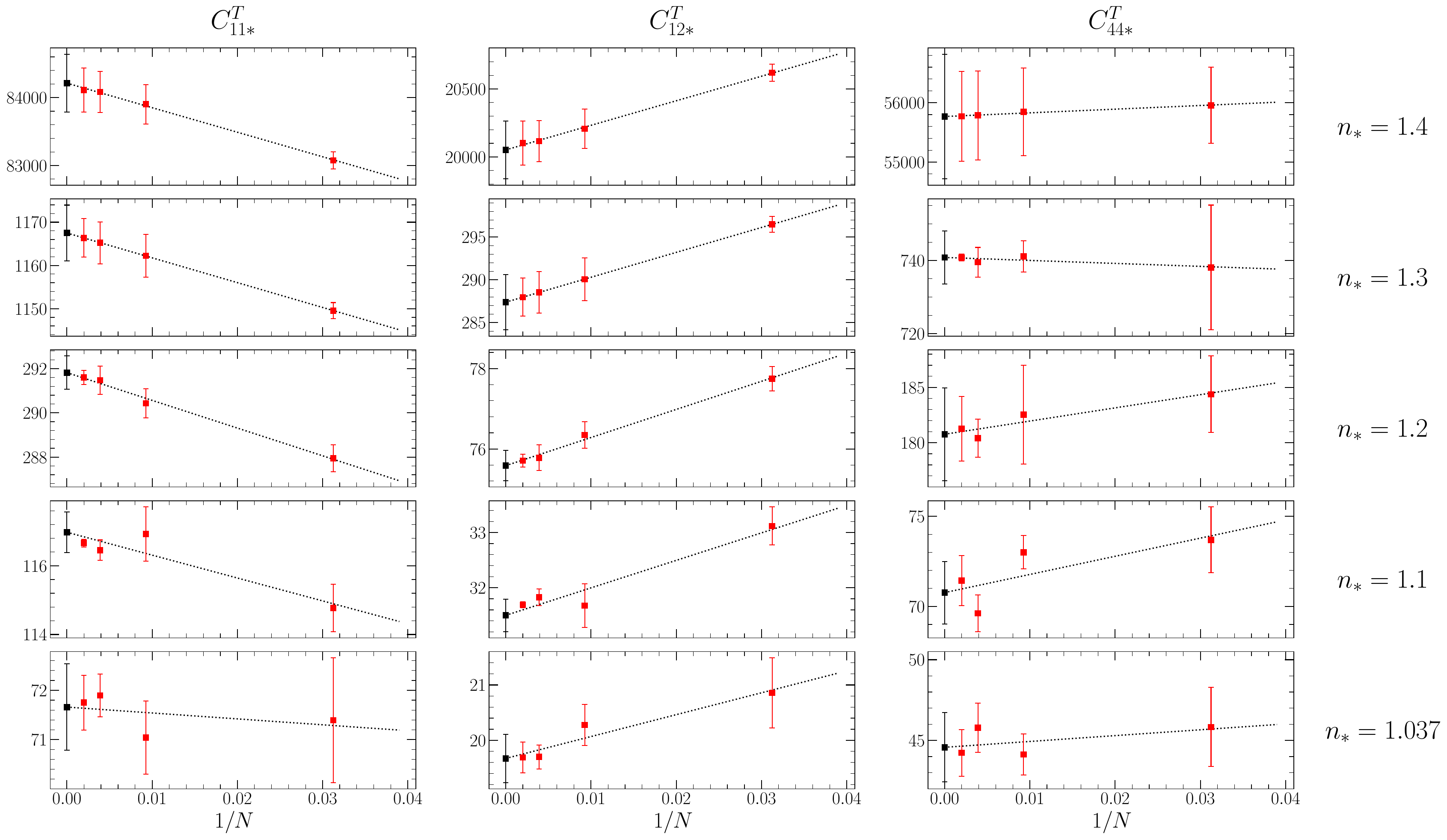}}
\caption[] {Isothermal elastic constants $C^T_{11}$, $C^T_{12}$, and $C^T_{44}$ (table~\ref{Tab:EC}) versus the dependence $N^{-1}$ on the particle number $N$ and the density $n_*$ (symbols) with the linear regression approximation (dotted lines) for the extrapolation $N\to\infty$.}\label{Fig:C111244}
\end{figure}

\begin{figure}[h!]\centering
{\includegraphics[width=1.\textwidth]{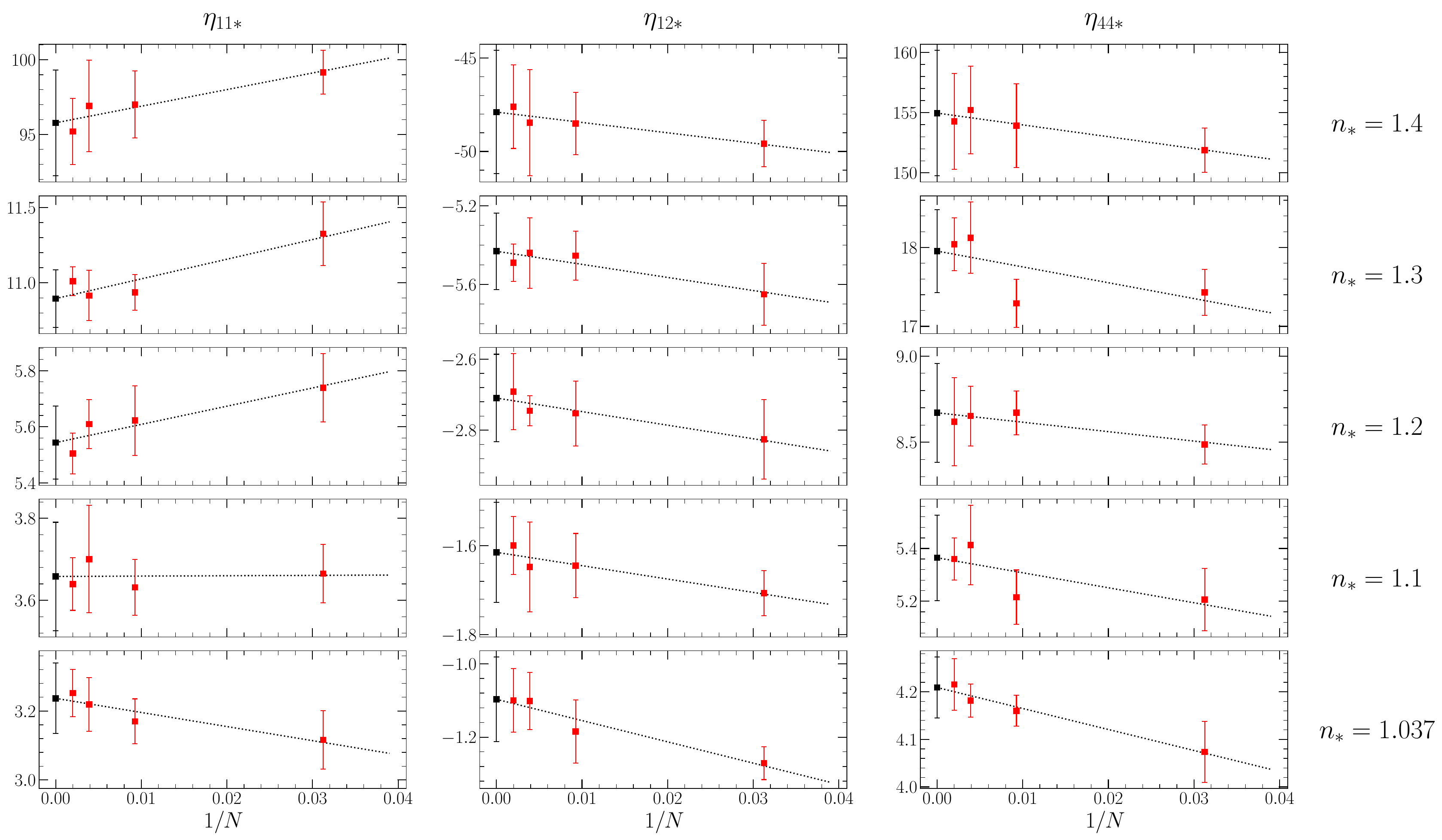}}
\caption[] {Viscosities $\eta_{11}$, $\eta_{12}$, and $\eta_{44}$  (table~\ref{Tab:eta}) versus the dependence $N^{-1}$ on the particle number $N$ and the density $n_*$ (symbols) with the linear regression approximation (dotted lines) for the extrapolation $N\to\infty$.}\label{Fig:eta111244}
\end{figure}

\begin{figure}[h!]\centering
{\includegraphics[width=.4\textwidth]{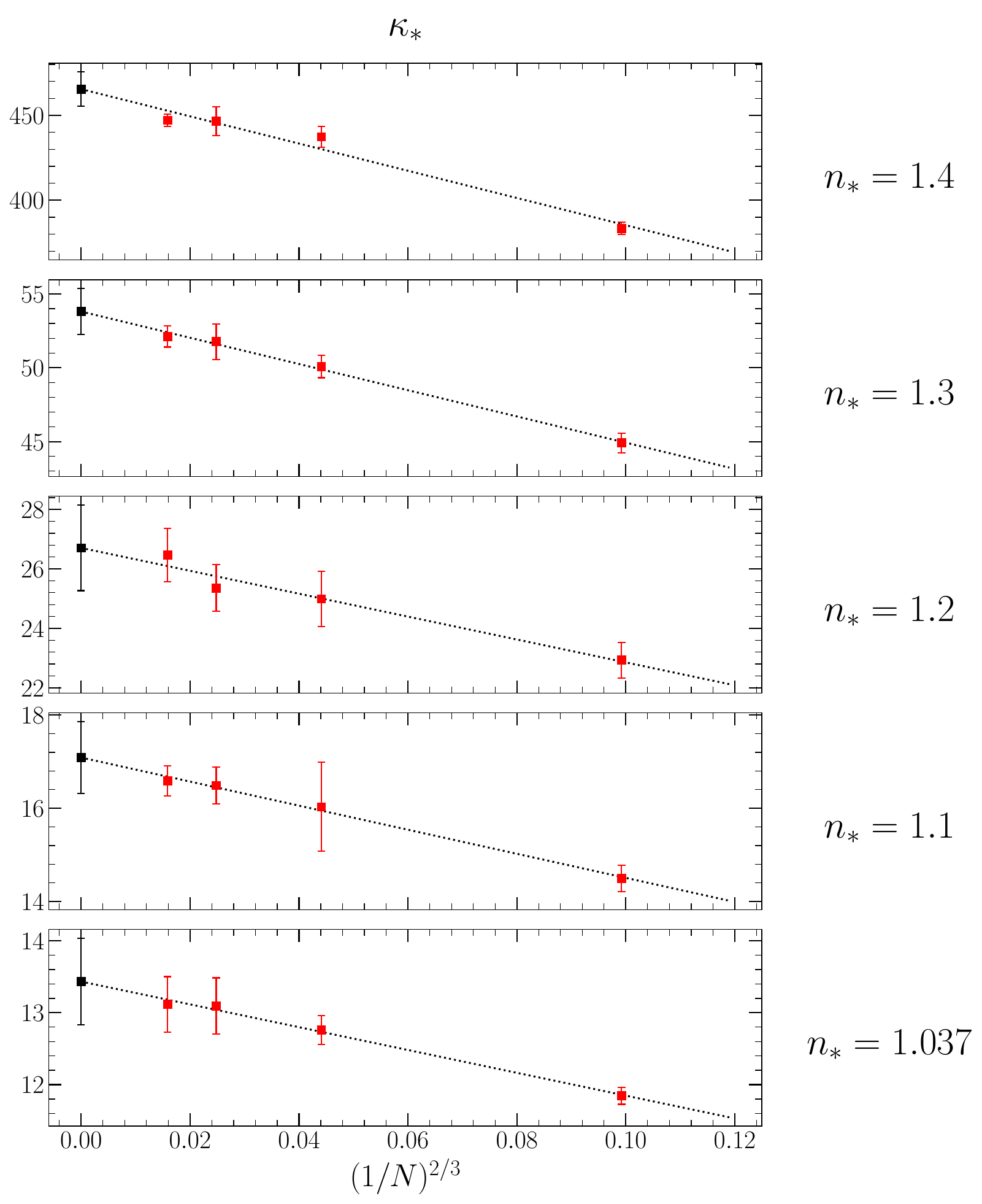}}
\caption[] {Heat conductivity $\kappa$ (table~\ref{Tab:kappa}) versus the dependence $N^{-2/3}$ on the particle number $N$ and the density $n_*$ (symbols) with the linear regression approximation (dotted lines) for the extrapolation $N\to\infty$.}\label{Fig:kappa}
\end{figure}

To estimate the values of the observables at infinite $N$, a  linear least squares fitting is performed on the data as function of $N^{-\alpha}$ with the exponent $\alpha=1$, except for the heat conductivity, for which the exponent $\alpha=2/3$ should be used as shown in reference~\cite{PBHB20}.
We consider data points $\{x_i,y_i\}_{i=1}^M$ with $x_i=N_i^{-\alpha}$, as shown in figures~\ref{Fig:pressure} for the  collision frequency and the pressure with $\alpha=1$, figure~\ref{Fig:C111244} for the elastic constants with $\alpha=1$, figure~\ref{Fig:eta111244} for the viscosities with $\alpha=1$, and figure~\ref{Fig:kappa} for the heat conductivity with $\alpha=2/3$.  There are fluctuations of the values $y_i$ denoted by $\Delta y_i$, but no fluctuations on the values $x_i$.  Linear least squares fitting is obtained by considering the minimum of
\begin{align}
R = \frac{1}{2} \sum_{i=1}^M (y_i-a-b x_i)^2 \, ,
\end{align}
so that the line $y=a+bx$ is fitted to the data with the slope $b$ and the ordinate at origin $a$.  These two coefficients are respectively given by the following well-known formulas:
\begin{align}
b &= \frac{\overline{x\, y}-\bar x\, \bar y}{\overline{x^2} - {\bar x}^2} \, ,\\
a &= \bar y -  \bar x \, b \, ,
\end{align}
where
\begin{align}
\overline{(\cdot)} \equiv \frac{1}{M} \, \sum_{i=1}^M (\cdot) \, .
\end{align}
 The error on the ordinate at origin is evaluated as 
 \begin{align}
\Delta a \simeq \sqrt{\frac{1}{M} \sum_{i=1}^M \left( \frac{ {\bar x}^2-\bar x \, x_i}{\overline{x^2} - {\bar x}^2}\right)^2 \Delta y_i ^2 \ } \, .
\end{align}

\section{Enskog theory for the heat conductivity}
\label{app:Enskog}

\begin{figure}[h!]\centering
{\includegraphics[width=.5\textwidth]{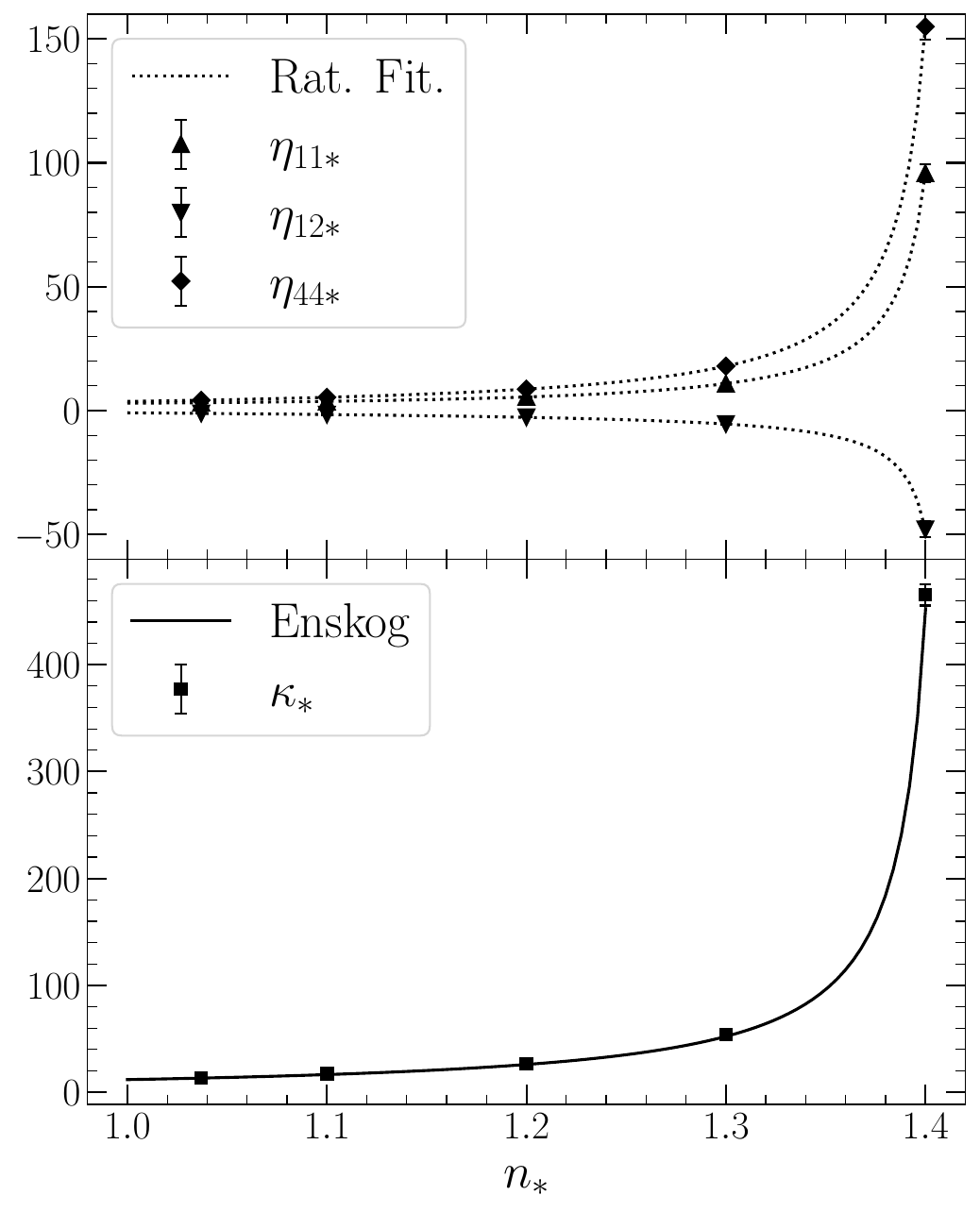}}
\caption[] {Transport coefficients versus the density $n_*$: The viscosities $\eta_{11}$, $\eta_{12}$, and $\eta_{44}$ are given in table~\ref{Tab:eta}, and the heat conductivity $\kappa$ in table~\ref{Tab:kappa}. The reported values are the extrapolation $N\rightarrow\infty$ from the results of the molecular dynamics simulation at various numbers of particles. When not appearing, the error bars are within the size of the symbols. As a guide of the eye, the dotted lines show the rational function~\eqref{eq:ratvisc} fitted to the data. The solid line shows the Enskog approximation~\eqref{eq:Enskogkappa} for the heat conductivity, using the equation of state~\eqref{eq:Ansatz_eos} and the parameters of reference~\cite{S98}.
}\label{Fig:TC}
\end{figure}

In the crystalline phase, the heat conductivity can be very well approximated by Enskog's kinetic theory \cite{RD77,DvBK21} according to
\begin{align}
\label{eq:Enskogkappa}
\kappa_{\rm E} &= B_2 n \left( \frac{1}{y_{\rm E}} + \frac{6}{5} +  0.757 \, y_{\rm E} \right) \kappa_{\rm B} \, , 
\end{align}
where $B_2=2\pi d^3/3$,  
\begin{align}\label{eq:y_E}
y_{\rm E} = \frac{p(n,T)}{n k_{\rm B} T} -1 = \frac{f(n)}{n}-1 \, ,
\end{align}
which is here evaluated with the truncated function~(\ref{eq:Ansatz_eos}), and
\begin{align}
\kappa_{\rm B} =(1+ 0.025 ) \frac{75 \, k_{\rm B}}{64 \, d^2} \, \sqrt{\frac{k_{\rm B}T}{\pi \, m}} \, ,
\end{align}
 which is the low-density approximation of Boltzmann's kinetic theory for the heat conductivity. These Enskog values for the heat conductivity are given in the last column of table~\ref{Tab:kappa} and shown in figure~\ref{Fig:TC} as a function of the particle density.




\pagebreak


\begin{table}[h!]
  \begin{tabular}{c @{\hskip 0.5cm} c @{\hskip 1cm} c @{\hskip 0.5cm} c @{\hskip 0.5cm} c @{\hskip 0.5cm} c}
    \hline\hline
      $n_*$ &   $N=32$     &    $N=108$  &  $N=256$   &   $N=500$ & $N\rightarrow \infty$ \\
    \hline  
    1.037	 & $33.550\pm0.006$  & $34.078\pm0.002$ &$34.204\pm0.003$ &$34.256\pm0.002$ & $34.301\pm0.004$\\   
    1.1 	& $40.764\pm0.005$  &$41.304\pm0.002$  & $41.442\pm0.002$ &$41.480\pm 0.001$ & $41.533\pm0.003$ \\ 
    1.2 	& $61.562\pm0.004$  & $62.237\pm0.003$ & $62.402\pm0.003$ & $62.450\pm0.002$ & $62.517\pm 0.003$ \\ 
    1.3	& $119.320\pm0.005$  & $120.532\pm0.005$ & $120.819\pm0.002$ & $120.920\pm0.002$& $121.033\pm0.004$ \\ 
    1.4  	& $992.37\pm0.01$  & $1001.68\pm0.01$  & $1003.93\pm0.04$ & $1004.77\pm0.04$ & $1005.60\pm0.05$ \\         
    \hline\hline
  \end{tabular}
\caption{Collision frequency per particle $\nu$ computed in the molecular dynamics simulations versus the particle number $N$ and the density $n_*$. The last column is the extrapolation $N\rightarrow\infty$ using the dependence $N^{-1}$.}\label{Tab:collfreq}
\end{table}


\begin{table}[h!]
  \begin{tabular}{c @{\hskip 0.5cm} c @{\hskip 0.5cm} c @{\hskip 0.5cm} c @{\hskip 0.5cm} c @{\hskip 0.5cm} c  @{\hskip 0.5cm} c  }
    \hline\hline
      $n_*$ &   $N=32$     &    $N=108$  &  $N=256$   &   $N=500$ & $N\rightarrow \infty$ &	Speedy \\
    \hline  
    1.037	 &  $11.451\pm   0.009	$&$  11.518\pm0.005	$&$11.533\pm0.002	$&$	11.539\pm0.001	$&$	11.545\pm0.005	$&$ 11.542$\\   
    1.1 	&    $14.528\pm0.005 	$&$	14.574\pm0.002	$&$14.586\pm0.001	$&$ 	14.590\pm0.001 	$&$  14.594\pm0.003	$&$  14.593$	\\ 
    1.2 	&   $23.315\pm0.007 	$&$	23.348\pm0.005	$&$	 23.357\pm0.003	$&$	23.359\pm0.001	$&$	23.362\pm0.004	$&$	 23.361$	\\ 
    1.3	&     $47.738\pm0.010	$&$	47.767\pm0.005	$&$ 47.773\pm0.005	$&$	47.776\pm0.004	$&$	 47.778\pm0.007	$&$	47.778$\\ 
    1.4  	&    $417.252\pm0.008	$&$	417.277\pm0.007	$&$ 417.283\pm0.017	$&$	 417.285\pm0.010	$&$	417.288\pm0.017	$&$ 417.287$ 	\\         
    \hline\hline
  \end{tabular}
  \caption{Pressure $p$ computed with equation~\eqref{eq:pressure} versus the particle number $N$ and the density $n_*$. The penultimate column is  the extrapolation $N\rightarrow\infty$ using the dependence $N^{-1}$. The last column gives the values of the equations of state~\eqref{eq:Ansatz_eos} with the parameters of reference~\cite{S98}. }\label{Tab:pressure}
\end{table}


\begin{table}[h!]
  \begin{tabular}{c @{\hskip 1cm}c @{\hskip .2cm}c@{\hskip 1cm}c@{\hskip .2cm}c @{\hskip 1cm}c@{\hskip .2cm}c}
    \hline\hline
    $n_*$ & $B_{T}$ &$B_{T*{\rm S}}$ & $c_{p*}$ & $c_{p*{\rm S}}$ & $\gamma$ & $\gamma_{\rm S}$   \\ 
    \hline
    1.037 & 40.837 & 40.874 & 4.647 & 4.643 & 3.098 & 3.095  \\
    1.1 & 64.856 & 64.868 & 4.485 & 4.484 & 2.990 & 2.990    \\
    1.2 &155.453 & 155.455 & 4.426 & 4.426 & 2.950 & 2.950    \\
    1.3 & 596.689 & 596.687 & 4.443 &  4.443 & 2.962 & 2.962   \\
    1.4  & 41577.763 & 41577.761 & 4.491 & 4.491 & 2.994 & 2.994  
     \\      \hline\hline
  \end{tabular}
  \caption{Thermodynamic properties  after the extrapolation $N\to\infty$ versus the density $n_*$:  The isothermal bulk modulus $B_T$ is given by equation~\eqref{eq:B_T}, the specific heat capacity $c_p$ by equation~\eqref{eq:c_P}, the specific heat ratio $\gamma$ by equation~\eqref{eq:gamma} as computed using molecular dynamics simulation and compared to the values predicted by equation of state~\eqref{eq:Ansatz_eos} using the parameters of reference~\cite{S98}. }\label{Tab:TH}
\end{table}


\begin{table}[h!]
  \begin{tabular}{l @{\hskip .2cm} c @{\hskip .5cm} c @{\hskip .5cm} c @{\hskip .5cm} c @{\hskip .5cm} c  @{\hskip .5cm} c }
    \hline\hline
     & $n_*$ &   $N=32$     &    $N=108$  &  $N=256$   &   $N=500$ &  $ N\rightarrow \infty$  \\
    \hline  
  & 1.037	 &  $  71.39 \pm 1.26	$&$ 71.04  \pm 0.74	$&$ 71.90 \pm 0.43	$&$ 71.75 \pm 	0.55$&	$ 71.66\pm0.88$\\   
   & 1.1 	&    $114.77 \pm 0.69	$&$	116.92\pm 0.79$&$ 116.46 \pm 0.30	$&$116.66 \pm 0.11 $& $ 116.98 \pm 0.59$ 	\\ 
   $C^T_{11*}$&   1.2	&$  287.95  \pm 0.60	$&$290.43	\pm 0.66$&$	291.48\pm 0.64 $&$291.61 \pm 0.32	$&$291.82 \pm 0.75 $ 	\\ 
   & 1.3	&     $1 149.55\pm1.85	$&$ 1 162.23	\pm 4.96$&$  1 165.23	\pm 4.83 $&$1 166.36 \pm 4.45	$&$1 167.51 \pm 6.46 $\\ 
   & 1.4  	&    $83 076\pm127	$&$	83 902 \pm289	$&$ 84 084 \pm 304	$& $84 112 \pm324$	&$ 84 213 \pm 427$	\\      
       \hline  
  & 1.037	 & $20.86\pm 0.63   	$&$ 20.28  \pm 0.37$&$19.70 \pm 0.22	$&$19.69 \pm 0.28	$&$19.67	\pm0.44 $\\   
   & 1.1 	&   $ 33.12 \pm  0.35	$&$31.68	\pm 0.39$&$31.83 \pm 0.15$&$ 31.69  \pm 0.06 $&$ 31.50  \pm $	0.30\\ 
  $C^T_{12*}$&   1.2	& $ 77.75 \pm 0.30  $&$76.35	\pm 0.33$&$ 75.78\pm 0.32  $&$75.71\pm  0.16	$&$75.58 \pm 0.38$	\\ 
   & 1.3	&     $296.48\pm 0.93	$&$	290.07\pm 2.48 $&$288.54 \pm 2.41	$&$287.97\pm 2.23	$&$ 287.39 \pm 3.23$\\ 
   & 1.4  	&    $20 620 \pm 64	$&$20 207	\pm 145 $&$ 20 116	\pm 152$& $20 102\pm  162	$	&$20 051\pm 214 $	\\         
       \hline  
 & 1.037 &$ 45.83\pm 2.45	 $&$  44.13  \pm1.27 $&$ 45.79  \pm  1.53	$&$ 44.23\pm 1.44	$&$44.58 \pm2.15 	$\\   
   & 1.1 	&$   73.69  \pm 1.83	$&$ 73.00	 \pm 0.92 $&$69.62 \pm 1.02	$&$  71.43 \pm1.39  $& $ 70.77 \pm 1.73	$\\ 
  $C^T_{44*}$ &  1.2 	&$  184.38  \pm 3.46	$&$	182.54 \pm 4.45$&$	180.42 \pm 1.73$&$181.27 \pm 2.93	$&$180.77 \pm 4.20$	\\ 
   & 1.3	& $  738.09  \pm 17.04	$&$ 741.10	\pm 4.30$&$ 739.51 \pm 4.05	$&$740.81 \pm 0.95	$& $740.82 \pm 7.22$\\ 
   & 1.4  	&  $55 957  \pm 	641 $&$	55 846 \pm 737$&$ 55 788 \pm 748	$& $55 770 \pm 758	$	& $ 55 769\pm 1051$	\\            
    \hline\hline
  \end{tabular}
  \caption{Isothermal elastic constants $C^T_{11}$, $C^T_{12}$, and $C^T_{44}$ computed with equations~\eqref{eq:C11mC12},~\eqref{eq:Cbulkm}, and~\eqref{eq:C44-zx} versus the particle number $N$ and the density $n_*$. The last column is  the extrapolation $N\rightarrow\infty$ using the dependence $N^{-1}$.}\label{Tab:EC}
\end{table}


\begin{table}[h!]
  \begin{tabular}{l @{\hskip .2cm} c @{\hskip .5cm} c @{\hskip .5cm} c @{\hskip .5cm} c @{\hskip .5cm} c  @{\hskip .5cm} c }
    \hline\hline
     & $n_*$ &   $N=32$     &    $N=108$  &  $N=256$   &   $N=500$ &  $ N\rightarrow \infty$  \\
    \hline  
  & 1.037	 &  $3.116   \pm 0.084	$&$   3.170\pm0.065	$&$3.219\pm 0.078	$&$3.252\pm 0.069	$&	$3.237\pm0.102$\\   
   & 1.1 	&    $3.665 \pm 0.071	$&$	3.631\pm 0.068$&$3.701\pm 0.131	$&$ 3.640\pm 0.064 $& $ 3.658\pm 0.132$ 	\\ 
   $\eta_{11*}$&  1.2 	&$   5.739 \pm 0.121	$&$	5.622\pm 0.124$&$	 5.610\pm 0.087$&$5.504\pm 0.073	$&$5.543\pm 0.130$ 	\\ 
   & 1.3	&     $11.325\pm 0.210	$&$	10.936\pm 0.117$&$ 10.915	\pm 0.167$&$11.010\pm 0.095	$&$ 10.895\pm 0.191$\\ 
   & 1.4  	&    $99.17\pm 1.47	$&$	97.01\pm 2.25	$&$ 96.91\pm  3.07	$& $ 95.21\pm  2.22	$	&$95.78\pm 3.55$	\\      
       \hline  
  & 1.037	 & $-1.271\pm 0.045  	$&$   -1.184\pm 0.086$&$-1.102\pm 0.078	$&$-1.100\pm 0.087	$&$	-1.097\pm 0.115$\\   
   & 1.1 	&   $ -1.707 \pm 0.051	$&$	-1.645\pm 0.072$&$-1.648\pm 0.101	$&$ -1.599 \pm  0.065$&$  -1.615\pm 0.113$	\\ 
  $\eta_{12*}$&  1.2 	& $  -2.826\pm 0.112 $&$	-2.753\pm 0.091$&$	 -2.745\pm  0.043$&$-2.692\pm  0.107	$&$-2.709\pm 0.123$	\\ 
   & 1.3	&     $-5.650\pm 0.156	$&$	-5.454\pm 0.125$&$ -5.440\pm 0.179	$&$-5.490\pm 0.095	$&$  -5.432\pm 0.195$\\ 
   & 1.4  	&    $-49.58\pm 1.24	$&$	-48.50\pm  1.67$&$ -48.46	\pm 2.85$& $ -47.60\pm  2.24	$	&$-47.89\pm 3.30 $	\\         
       \hline  
 & 1.037 &$ 4.074\pm 0.064	 $&$  4.160   \pm 0.033	$&$  4.182\pm  0.035	$&$ 4.216\pm 0.054	$&$4.209\pm 0.064	$\\   
   & 1.1 	&$    5.207 \pm 0.118	$&$	5.216 \pm  0.103$&$5.413\pm 0.150	$&$ 5.360\pm 0.080 $& $  5.365\pm 0.162	$\\ 
  $\eta_{44*}$ &  1.2 	&$   8.487 \pm 0.113	$&$	8.671\pm 0.127$&$	 8.653\pm 0.173$&$8.619\pm0.256	$&$8.671\pm 0.287$	\\ 
   & 1.3	& $    17.429\pm 0.292	$&$	17.292\pm 0.306$&$ 18.123 \pm 0.453	$&$18.041 \pm 0.334	$& $17.955	\pm0.526$\\ 
   & 1.4  	&  $  151.89\pm 1.83	$&$	153.91\pm 3.48$&$ 155.22\pm 3.65	$& $ 154.27\pm  3.99	$	& $154.96\pm 5.20$	\\            
    \hline\hline
  \end{tabular}
  \caption{Viscosities $\eta_{11}$, $\eta_{12}$, and $\eta_{44}$ computed with equation~\eqref{eq:etaH} versus the particle number $N$ and the density $n_*$. The last column is  the extrapolation $N\rightarrow\infty$ using the dependence $N^{-1}$.}\label{Tab:eta}
\end{table}


\maketitle
\begin{table}[h!]
  \begin{tabular}{c @{\hskip 1cm} c @{\hskip 1cm} c @{\hskip 1cm} c @{\hskip 1cm} c @{\hskip 1cm} c @{\hskip 1cm} c   }
    \hline\hline
      $n_*$ &   $N=32$     &    $N=108$  &  $N=256$   &   $N=500$ & $N\rightarrow \infty$ & Enskog  \\
    \hline  
    1.037	 &$  11.845  \pm 0.118 	$&$  12.762	\pm 0.200$&$13.093\pm 0.393	$&$13.117\pm 0.385	$&$ 13.434\pm  0.601$ &13.199\\   
    1.1 	&   $ 14.496 \pm 0.285	$&$	16.029\pm  0.955$&$16.490\pm 0.391	$&$  16.588\pm0.324 $&  $17.088\pm 0.767$ &16.498	\\ 
    1.2 	&   $22.93 \pm 0.60	$&$	24.99\pm 0.93$&$	 25.36\pm 0.78$&$26.46\pm 0.90	$&$26.71\pm1.44$&	25.95\\ 
    1.3	&    $ 44.91	\pm 0.66$&$	50.08\pm 0.75$&$ 51.77\pm 1.21	$&$52.12\pm 0.72	$& $53.81\pm1.55$&52.20\\ 
    1.4  	&   $ 383.41\pm 3.57	$&$	437.31\pm 6.16	$&$ 446.52\pm 8.48	$&	$447.08\pm 3.60$&$465.4\pm 10.06$& 449.24\\         
    \hline\hline
  \end{tabular}
  \caption{Heat conductivity $\kappa$ computed with equation~\eqref{eq:kappaH} versus the particle number $N$ and the density $n_*$. The penultimate column is the extrapolation $N\rightarrow\infty$ using the dependence $N^{-2/3}$. The last column gives the values of the Enskog approximation~\eqref{eq:Enskogkappa} with the equation of state~\eqref{eq:Ansatz_eos} and the parameters of reference~\cite{S98}. }\label{Tab:kappa}
\end{table}

\pagebreak

\begin{figure}[h!]\centering
{\includegraphics[width=.7\textwidth]{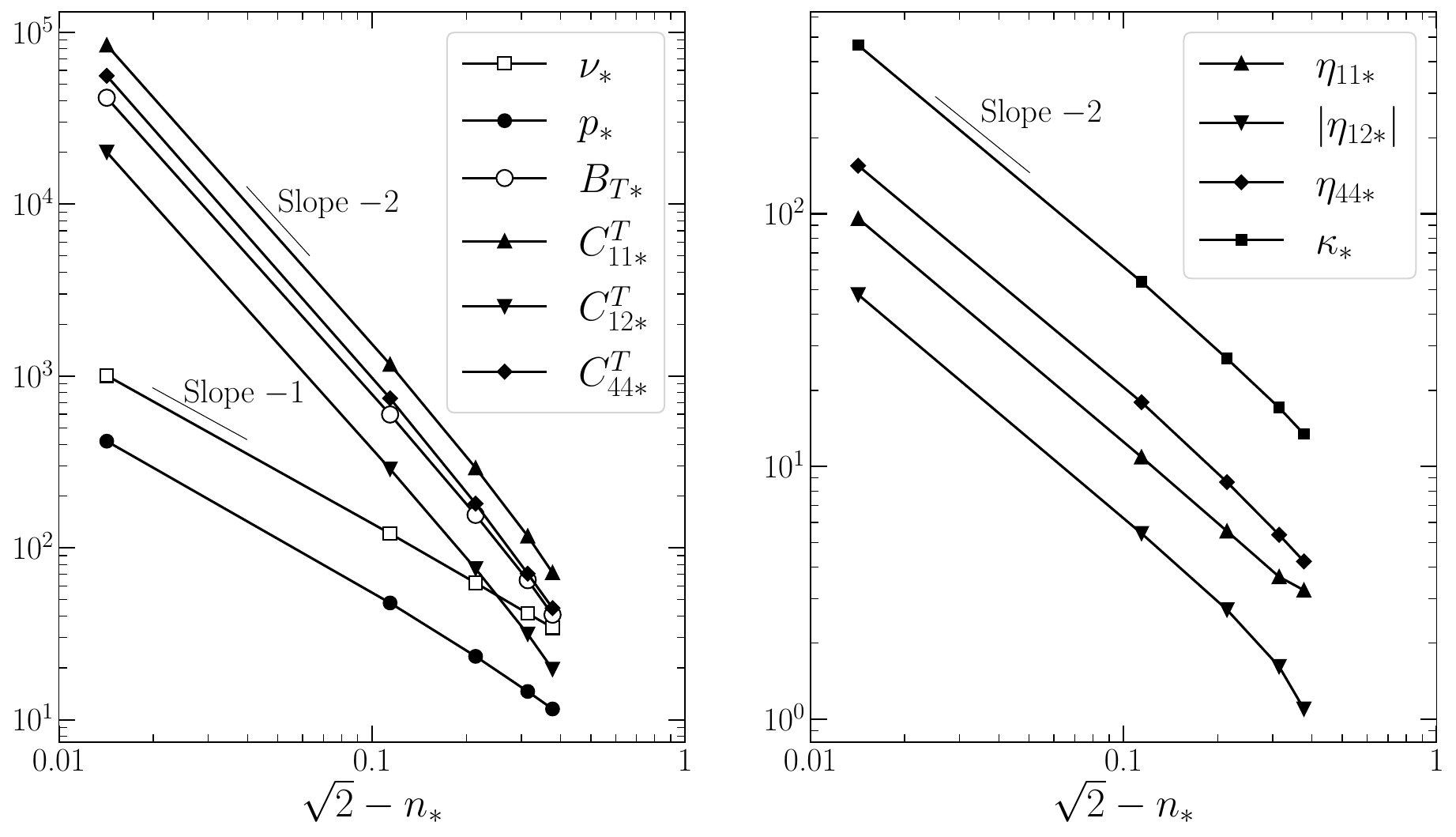}}
\caption[] {Divergences of the equilibrium and nonequilibrium hydrodynamic properties in the vicinity of the close-packing density $n_*=\sqrt{2}$.  Left: Log-log plot of the collision frequency per particle $\nu$ (table~\ref{Tab:collfreq}), the hydrostatic pressure $p$ (table~\ref{Tab:pressure}), the isothermal bulk modulus $B_T$ (table~\ref{Tab:TH}), and the elastic constants $(C_{11}^T,C_{12}^T,C_{44}^T)$ (table~\ref{Tab:EC}) versus $\sqrt{2}-n_*$. Right: Log-log plot of the viscosities $(\eta_{11},\eta_{12},\eta_{44})$ (table~\ref{Tab:eta}) and the heat conductivity $\kappa$ (table~\ref{Tab:kappa}) versus $\sqrt{2}-n_*$.}\label{Fig:divergences}
\end{figure}


\begin{figure}[h!]\centering
{\includegraphics[width=1.0\textwidth]{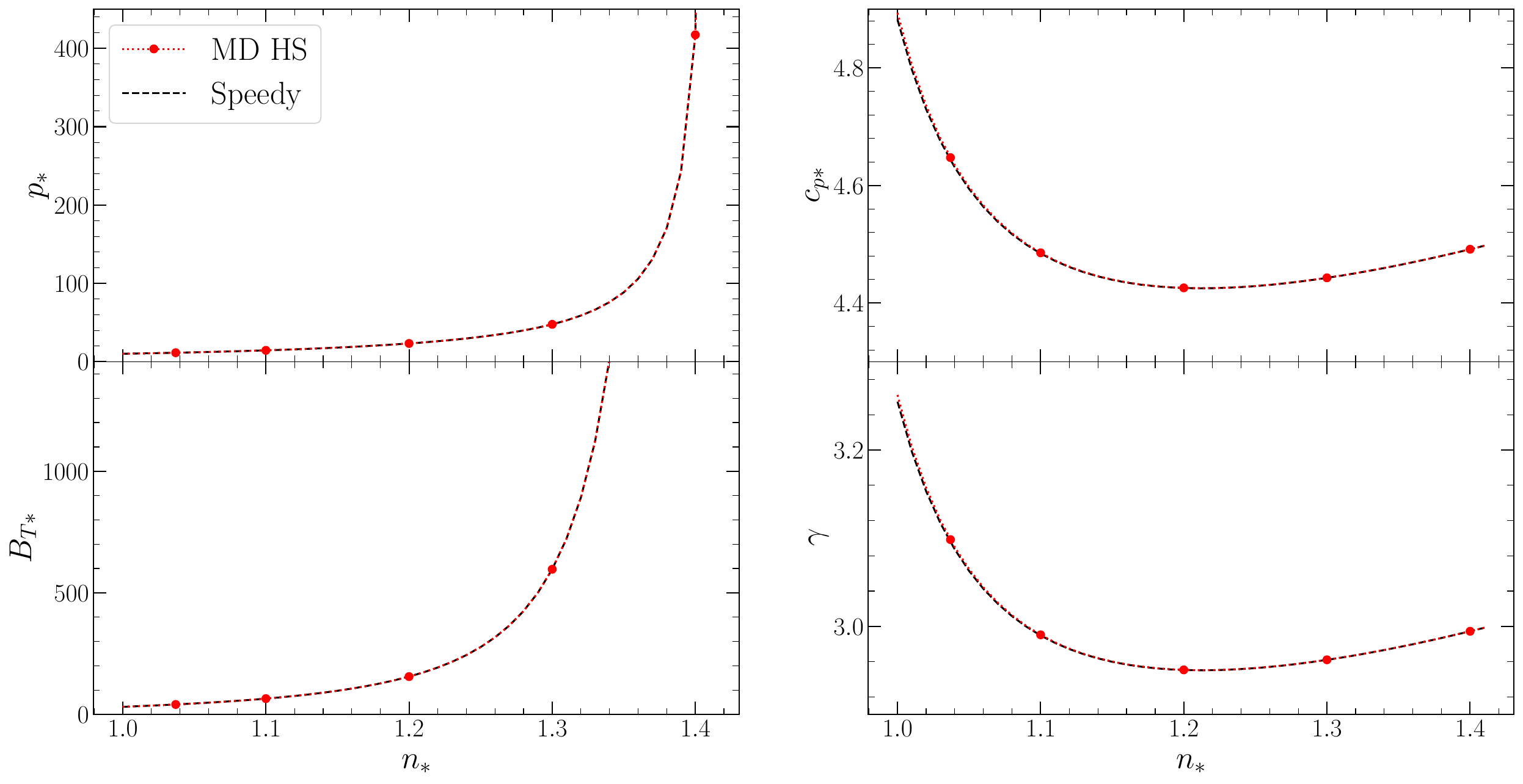}}
\caption[] {Thermodynamic properties  after the extrapolation $N\to\infty$ versus the density $n_*$:  The pressure $p$ is given in table~\ref{Tab:pressure}, and the isothermal bulk modulus $B_T$, the specific heat capacity $c_p$, and the specific heat ratio $\gamma$ in table~\ref{Tab:TH}. The results computed using molecular dynamics simulation (dots) are compared to the values predicted by equation of state~\eqref{eq:Ansatz_eos} using the parameters of reference~\cite{S98} (dashed lines).}\label{Fig:TH}
\end{figure}


\begin{figure}[h!]\centering
{\includegraphics[width=.5\textwidth]{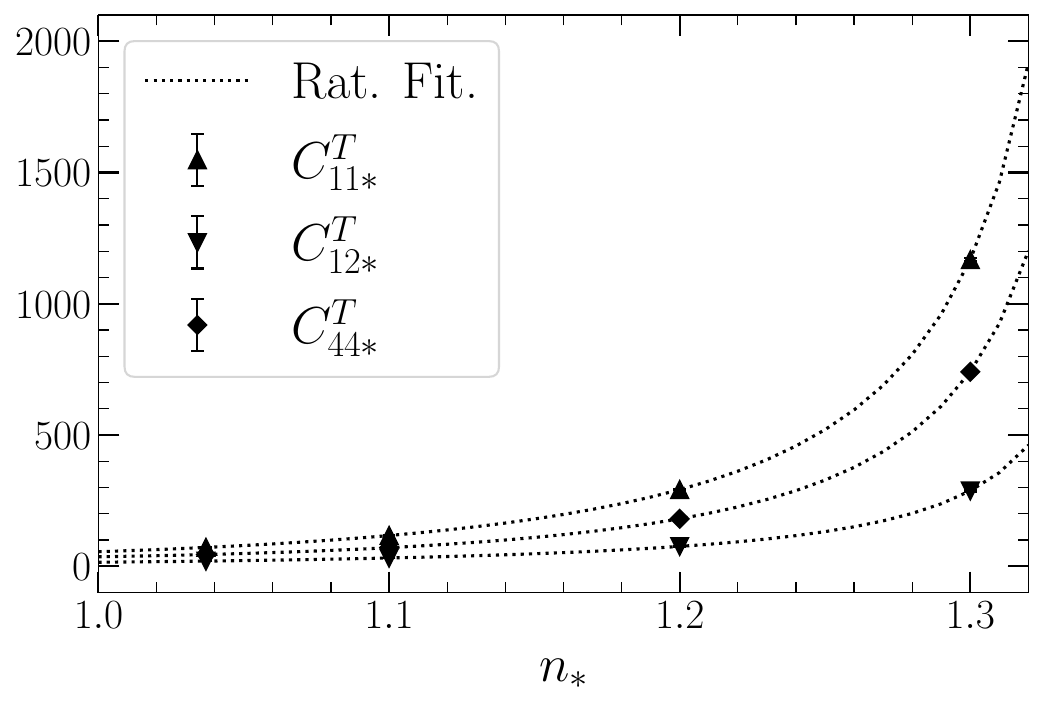}}
\caption[] {Isothermal elastic constants after the extrapolation $N\to\infty$ given in table~\ref{Tab:EC} versus the density $n_*$. As a guide of the eye, the dotted lines show the rational function~\eqref{eq:ratvisc} fitted to the data. The values at $n_*=1.4$ are not shown because they are very large.}\label{Fig:EC}
\end{figure}


\begin{thebibliography}{99}
\bibitem{footnote}  We note that more accurate values have been obtained in reference~\cite{PBBCH19} for the freezing density $n_{*{\rm f}}=0.9392\pm 0.0001$, the melting density $n_{*{\rm m}}=1.0376\pm 0.0001$, and the fluid-crystal coexistence pressure $p_{*{\rm FC}}=11.5712 \pm 0.0010$.
\end{thebibliography}
\end{document}